\documentclass[useAMS,usenatbib]{mn2e}

\usepackage{amsmath,amssymb,graphicx}

\newcommand{\kepler}{\textit{Kepler}}

\newcommand{\prat}{\mathcal{P}}

\title[Kepler Period Ratio Distribution]{The period ratio distribution of \kepler 's candidate multiplanet systems}
\author[Steffen \& Hwang]{
Jason H. Steffen$^{1,2}$ and Jason A. Hwang$^{1}$
\\
$^{1}$CIERA, Northwestern University, 2145 Sheridan Road, Evanston, IL 60208\\
$^{2}$Lindheimer Fellow\\
}

\begin{document}


\pagerange{\pageref{firstpage}--\pageref{lastpage}} 

\maketitle

\label{firstpage}

\begin{abstract}
We calculate and analyze the distribution of period ratios observed in systems of \kepler\ exoplanet candidates including studies of both adjacent planet pairs and all planet pairs.  These distributions account for both the geometrical bias against detecting more distant planets and the effects of incompleteness due to planets missed by the data reduction pipeline.  In addition to some of the known features near first-order mean-motion resonances (MMR), there is a significant excess of planet pairs with period ratios near 2.2.  The statistical significance of this feature is assessed using Monte Carlo simulation.  We also investigate the distribution of period ratios near first-order MMR and compare different quantities used to measure this distribution.  We find that beyond period ratios of $\sim 2.5$, the distribution of all period ratios follows a power-law with an exponent ${-1.26 \pm 0.05}$.  We discuss implications that these results may have on the formation and dynamical evolution of \kepler -like planetary systems---systems of sub-Neptune/super-Earth planets with relatively short orbital periods.
\end{abstract}


\section{Introduction}

The architectures of planetary systems---the orbital properties and masses of the planets---are important observables for understanding the formation and dynamical evolution of those systems.  For example, the proximity of a pair of planets to Mean Motion Resonance (MMR) can be a strong indicator of past convergent migration between those planets (e.g., \citet{Lee:2002}).  Similarly, a significant lack of planet pairs near some period ratios would require a physical explanation.  The insights gained by identifying and explaining notable features in the architectures of planetary systems refine our knowledge of the histories of those systems, uncover the variety of dynamical paths that they may take, and improve our understanding of our own solar system in the context of the global population of planetary systems.

NASA's \kepler\ mission \citep{Borucki:2010} has proven exceptionally valuable for studying the architectures of the inner regions of planetary systems.  We know, for example, that \kepler\ observations favor systems with a relatively small number of planets whose orbits are coplanar to within a few degrees (\citet{Lissauer:2011b,Fabrycky:2014,Fang:2012,Tremaine:2012,Figueira:2012}---these last two references include RV data to support this statement) indicating that planets generally form within gaseous disks.  In addition, a large fraction of systems of \kepler -like planets are dynamically full \citep{Fang:2013}, meaning that additional, intermediate planets render the system unstable.

Studies of the distribution of orbital periods and period ratios in \kepler\ systems have also turned up a number of interesting features.  Planets in multiplanet systems that have very short orbital periods $\lesssim 2$ days are more isolated than planets with longer periods \citep{Steffen:2013c}.  The fraction of planet pairs near different MMRs is different for three-planet systems than it is for systems with four or more planets \citep{Steffen:2013b}.  There is a significant lack of planet pairs just interior to first-order MMRs, most strikingly near the 2:1 MMR \citep{Lissauer:2011b,Fabrycky:2014} and several studies have proposed an explanation for this feature involving various manifestations of processes such as tidal dissipation or interactions with gas or planetesimal disks \citep{Lithwick:2012a,Rein:2012,Batygin:2013,Lee:2013,Petrovich:2013,Chatterjee:2014,
Goldreich:2014,Xie:2014}.

Ultimately, the architectures of planetary systems, including the distribution of period ratios are key observations that inform planet formation models.  In-situ formation of planetary systems and systems assembled through the migration of planets from more distant regions generally predict different final architectures especially with regards to the proximity of planets to MMR \citep{Terquem:2007,Ida:2010,Bromley:2011,Baruteau:2013b,Chiang:2013,
Hansen:2013,Cossou:2014,Izidoro:2014,Raymond:2014,Schlaufman:2014}.  In this manuscript we use the most recent published catalog of \kepler\ data \citep{Burke:2014} to estimate the period ratio distribution of planets in the inner regions of planetary systems where \kepler\ discoveries concentrate.  Our aim is to identify significant, and hopefully useful, features to refine our understanding of how and where planets form, and how their orbits evolve dynamically through the various stages of the system lifetime.

We begin with a discussion of the corrections we make to the observed distribution of Kepler Objects of Interest (KOI) that account for mutual orbital inclinations of the planets, which causes some planets to not be observed, and for planets that are somehow missed by the \kepler\ data reduction pipeline.  Section \ref{adjacentplanets} presents and discusses the distribution of period ratios for adjacent planets while Section \ref{allplanets} has a similar investigation into the distribution of all planet pairs.  That section also quantifies the broad characteristics of the period ratio distribution.  In Section \ref{twopointtwo} we discuss two features from this analysis that appear to be statistically significant, but that have not been discussed at length elsewhere---namely, peaks near period ratios of 2.2 and 3.9.  Section \ref{firstorders} examines the planet pairs that are near first-order ($j+1$:$j$) MMR, showing how they are distributed based upon the index $j$ and what can be learned from the various ways to measure the ``distance'' from MMR.  We briefly consider the effects of including more recent \kepler\ data in Section \ref{quarter16} and finally we investigate the small set of systems with planet pairs near integer period ratios $j$:1 in Section \ref{integerratios}.  Our conclusions are summarized in Section \ref{conclusions}.

\section{Corrections for geometric bias and completeness}

In order to properly quantify the features in the period ratio distribution of the \kepler\ planet candidate sample, we must correct for two primary sources of bias---the geometric probability of a planet to transit the host star and the probability of a planet being identified by the \kepler\ pipeline.  We briefly outline our approach to each of these issues here and provide more details in the appendices.

Ultimately we consider the observed distribution of period ratios to equal
\begin{equation}\label{truedistribution}
D_\text{obs}(\prat) \propto \beta_\text{geo} \beta_\text{pipe} D_\text{true}(\prat)
\end{equation}
where $\prat = P_2/P_1$ is the ratio of orbital periods of the outer and inner planets, $\beta_\text{geo}$ is the bias due to geometry, and $\beta_\text{pipe}$ is the bias due to pipeline completeness (ultimately there is a normalization constant to preserve unit area).  Both $\beta$'s depend upon parameters such as the planet and stellar sizes, orbital distances, assumed distribution in mutual orbital inclinations, and the period ratio.  Fundamentally, however, each $\beta$ takes the form
\begin{equation}\label{betadef}
\beta = \frac{P_2(\prat)}{P_2(1)}
\end{equation}
where $P_2(x)$ is the probability of detecting the outer planet at a period ratio of $x$ with respect to the inner planet, given that the inner planet is known to transit the star.  Thus, the outer planet in a hypothetical co-orbital configuration serves as a fiducial value from which to calculate the relative probability of detecting the outer planet in its observed location.

\subsection{Geometric bias}

In this study our primary interest is the distribution of period ratios.  Thus, when we discuss the geometric bias, we do not consider planet size (it does play a role in the pipeline completeness calculation of the next subsection).  Instead, we consider the relative probability of a second planet transiting as a function of the distance in stellar radii of the inner planet from the host star, $a_1/R_\star$, and the ratio of the orbital periods of the planet pair, $P_2/P_1$.

\begin{figure}
\includegraphics[width=0.45\textwidth]{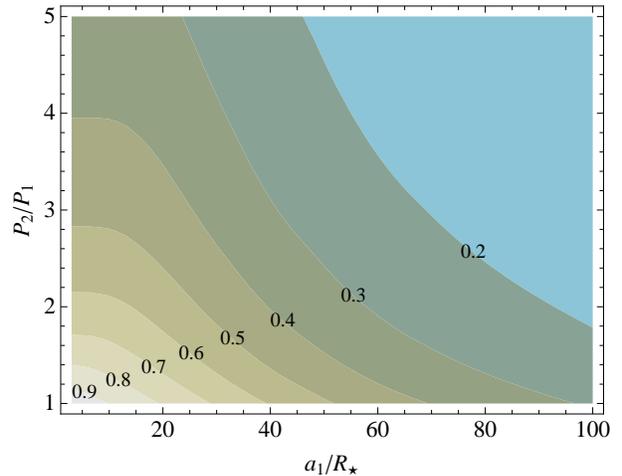}
\caption{Contours of constant probability for detecting an outer planet in a system given that an inner planet transits as a function of the period ratio of the two planets and the distance from the star to the inner planet in units of the stellar radius.  We assume that mutual orbital inclinations are Rayleigh distributed with a Rayleigh parameter of $1.5^{\circ}$.  These results are used to correct for the geometric bias when we calculate the orbital period ratio distribution.\label{geometry}}
\end{figure}

The details of this calculation are given in Appendix \ref{apxgeometry}.  The results of the correction are shown in Figure \ref{geometry}.  This figure shows the probability of detecting the outer planet planet in a pair---given that the inner planet has already been seen---as a function of the distance of the inner planet from the host star and the period ratio of the inner and outer planets.  For this figure we assume that the mutual inclinations of planetary orbits are Rayleigh distributed with a Rayleigh parameter (or ``width'') equal to 1.5$^\circ$.  This quantity is roughly consistent with estimates of the typical mutual inclination of \kepler\ systems \citep{Fang:2012,Lissauer:2011b,Tremaine:2012} and of the solar system planets.

\subsection{Pipeline completeness}

A completeness test of the \kepler\ pipeline, using the injection and recovery of synthetic transit signatures, is underway \citep{Christiansen:2013} but remains to be finished.  However, the TERRA pipeline has been tested in this manner \citep{Petigura:2012,Petigura:2013b}.  While there are differences between these two pipelines, particularly with regards to target selection and how they treat multiplanet systems (that is, the \kepler\ pipeline searches for them while TERRA does not), their results are largely the same for an individual planet detection.  Moreover, since we use the pipeline completeness results to determine the relative probability of detecting individual planets with specific sizes and orbital periods (thereby mitigating some of the differences between expected results), the TERRA esimates are suitable for our purposes.

\begin{figure}
\includegraphics[width=0.45\textwidth]{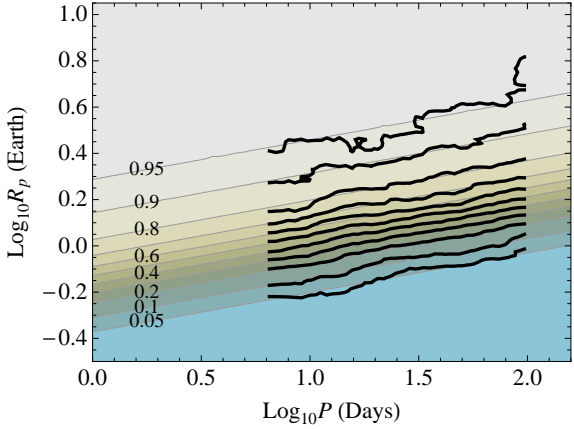}
\caption{Contours of constant pipeline detection completeness.  The thick lines are portions of contours from the data in \citet{Petigura:2013b}.  These selections were used to generate the straight, thin lines, which we employ in our calculations to correct for pipeline completeness.\label{completeness}}
\end{figure}

The details of the correction calculation are found in Appendix \ref{apxpipecomplete}.  The results are shown in Figure \ref{completeness} which has the detection probability as a funciton of planet orbital period and size.  The semi-analytic treatment that we employ is shown as the straight lines of constant detection probability.  The portion of the data from \citet{Petigura:2013b} that were used to derive this relationship are shown as the thick, non-straight lines.

\subsection{Combined corrections}

To produce our estimate for the distribution of orbital period ratios in \kepler\ planetary systems we first select all multiplanet systems from the \kepler\ planet candidate list given by \citet{Burke:2014} that do not have a disposition of ``False Positive''.  Systems that have a planet pair with period ratios less than 1.1 are excluded (e.g., KOI 284, see \citet{Lissauer:2014}) as are systems containing a planet larger than 20 $R_\oplus$ or where the largest distance from the host star to to a planet is larger than 250 $R_\star$ (this latter restriction affects two systems which we discuss in the next subsection).  For this work we study both the period ratio distribution of adjacent planet pairs and the period ratio distribution of all planet pairs.

We construct a kernel density estimator (KDE) of the distribution using a Gaussian kernel.  Specifically, for each measured period ratio we add a Gaussian-shaped distribution with a mean value equal to the measured period ratio and with a width (the ``bandwidth'') assigned based upon some criteria.  The corrections for the geometric transit probability and the pipeline completeness are applied to the KDE by appropriately increasing the area of the Gaussian kernel based upon the properties of that particular planet pair.

For our nominal distribution, the kernel bandwidth is equal to 0.005 times the measured period ratio.  For example, at a period ratio of 2:1, the bandwidth is 0.01 and near the 3:1 the bandwidth is 0.015.  The value of 0.005 was chosen so that the kernel of a planet pair near the 7:6 MMR (i.e., the period ratio near that of Kepler-36 \citet{Carter:2012}) would be well isolated from that for a planet pair near the 6:5.  We chose not to use a nearest-neighbor criteria for the bandwidth because there are relatively few planets with small period ratios, but the precise values of those period ratios can have important dynamical ramifications---a broad bandwith near high-index first-order MMRs would not be appropriate.

\begin{figure}
\includegraphics[width=0.45\textwidth]{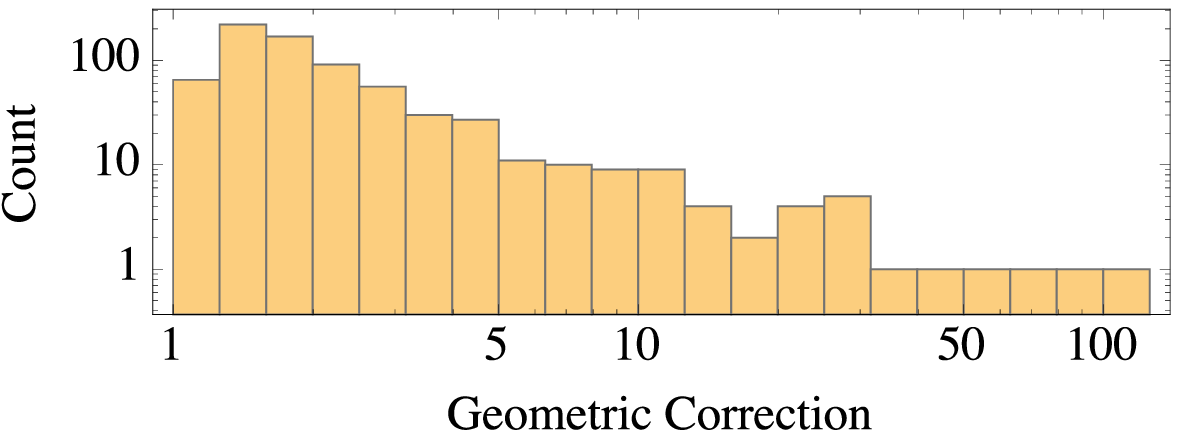}

\medskip
\includegraphics[width=0.45\textwidth]{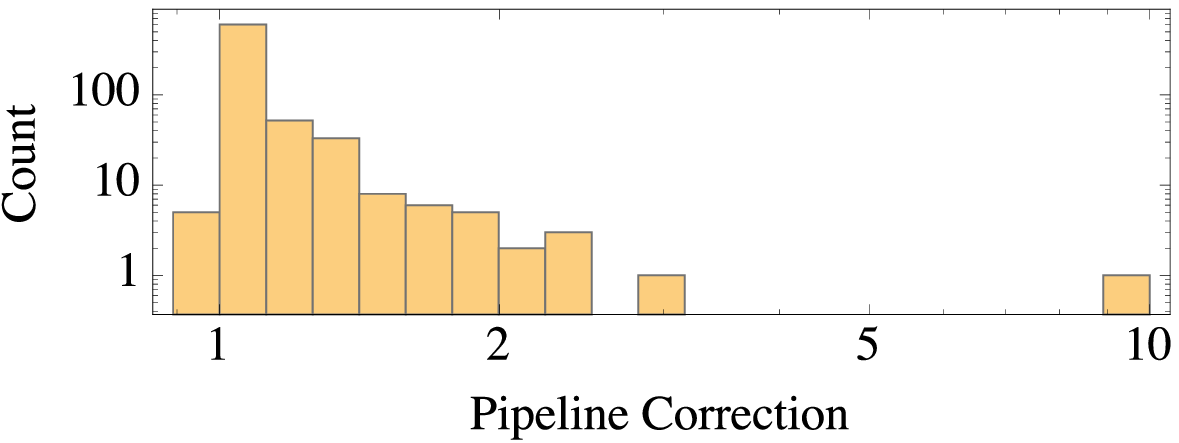}

\medskip
\includegraphics[width=0.45\textwidth]{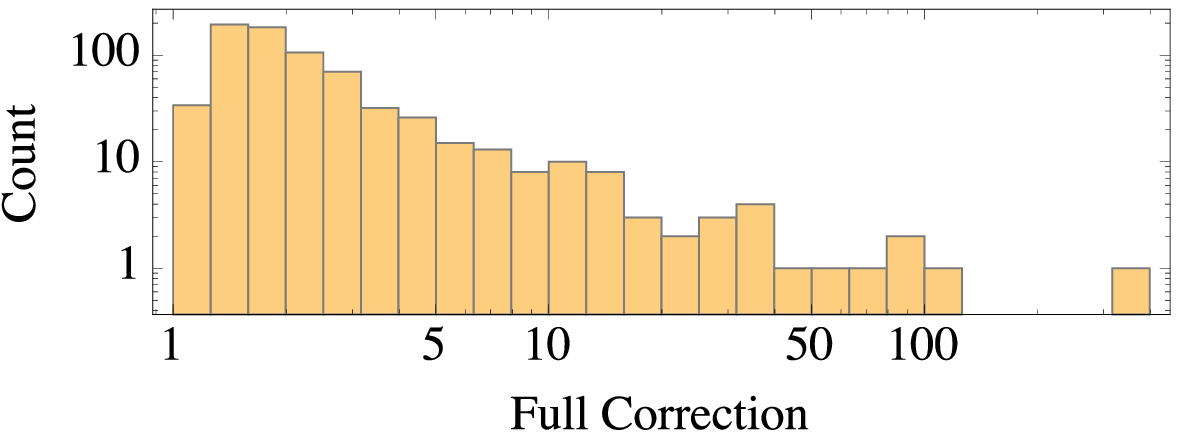}
\caption{The distribution of geometric corrections (top), pipeline completeness corrections (middle), and the total corrections (bottom) for adjacent planet pairs.  We note that five of the systems have a pipeline completeness correction that is slightly less than unity (typically the differences are $<0.001$ with the largest being 0.003).  These are due to small dips in the interpolation used to generate the pipeline completeness correction (see Appendix \ref{apxpipecomplete}).  Their effect on our results is negligible.\label{corrections}}
\end{figure}

Beginning with unit area, we increase the area of the kernel by dividing by the two $\beta$'s in Equations (\ref{truedistribution}) and (\ref{betadef}).  That is, we calculate the probability of detecting the outer planet at the location of the inner planet relative to the probability of detecting the outer planet at its observed location using the results shown in Figures \ref{geometry} and \ref{completeness} and increase the area of the kernel appropriately.  Figure \ref{corrections} shows how these two corrections and their combination are distributed for all planet pairs (for this figure we make no restriction on $a/R_\star$).

\begin{figure}
\includegraphics[width=0.45\textwidth]{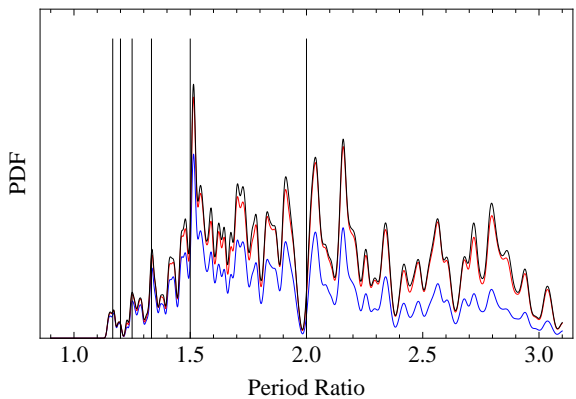}
\caption{Period ratio distribution from adjacent planet pairs showing the observed distribution (blue, lower curve) the effect of correcting for geometric bias (red, middle curve) and the effect of correcting for both the geometric and the pipeline completeness (black, top curve).  The top, black curve is our estimate for the true period ratio distribution.  Thin vertical lines mark the locations of first-order mean-motion resonances.\label{pdistshort}}
\end{figure}

The geometric correction tends to make the largest contribution with values of several tens being common.  The pipeline completeness correction makes changes closer to unity, with one notable exception discussed below.  An example distribution that results from these corrections is shown in Figure \ref{pdistshort}.  That figure shows the raw (observed) distribution along with the effects of the geometric and pipeline corrections.  We note that the pipeline completeness is a weak function of orbital period and makes very little difference.  Also shown in Figure \ref{pdistshort} are the locations of the first order MMRs from the 2:1 through the 7:6.

\subsection{KOIs 2311 and 435}\label{oddsandends}

As seen in the bottom panel of Figure \ref{corrections}, our approach can often alter a system's relative probability by factors of several tens.  KOI-2311 is unique in that its overall correction is nearly a factor of 400.  The primary cause of this large correction is the small sizes of the constituent planets.  With a relatively large period ratio of 13.979 (quite near 14:1) and moderately large orbital periods of 13.7 days and 192 days, its geometric correction is a factor of 44.15---already a significant quantity.  But, the planet sizes are 0.77 and 0.95 $R_\oplus$, which coupled with the large orbital period of the outer planet, places the system far in the tail of of the distribution of pipeline corrections.

Planets this small on orbits this large are difficult to detect and the fact that one has been seen indicates that Earth-sized planets on orbits of a few hundred days should be (and indeed are) quite common \citep{Petigura:2013b,Fressin:2013}.  However, since our primary goal is to study small-scale features of the period ratio distribution (using a relatively narrow smoothing kernel), including this system produces an exceptionally large spike near a period ratio of 14:1---ostensibly making it the most common period ratio in planetary systems (which it almost certainly is not).  It is this system that motivates our cut at $a/R_\star < 250$---the value of this quantity for the outer planet in KOI-2311 is 276.

The only other system excluded by this cut is KOI-435.  The $a/R_\star$ for the outer planet in this system is 291 and the planet orbital periods are 20.5 and 740 days (yielding a period ratio near 36).  However, the planets in this system are quite large at 3.6 and 8.8 $R_\oplus$ for the inner and outer planets respectively, which produces a pipeline completeness correction of only 1\%.  The geometric correction for this system is 71.  While, the overall effect of this system on the period ratio distribution is benign, it does not meet our chosen criteria and is not used in any of our analyses below.

\section{Adjacent pairs}\label{adjacentplanets}

\begin{figure}
\includegraphics[width=0.45\textwidth]{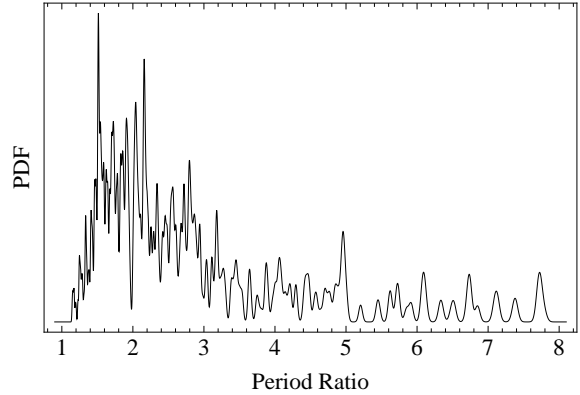}
\caption{The period ratio distribution for \kepler\ planets using only adjacent planet pairs and correcting for geometric bias and pipeline completeness.  The most prominent feature is the spike near the 3:2 MMR.  Second to that is the excess of planet pairs near a period ratio of 2.2.\label{pdist}}
\end{figure}

There are 474 multiplanet systems that satisfy our selection criteria that the smallest period ratio be larger than 1.1---to avoid unstable or split multiplanet systems (systems with planets orbiting two different stars), that the largest planet be smaller than 20 $R_\oplus$, and that the maximum value of $a/r_\star$ be less than 250.  Our criteria yields a total of 718 adjacent planet pairs from which to construct the period ratio distribution shown in Figure \ref{pdist}.

The broad features of the period ratio distribution include a sharp rise with a broad, overall peak between period ratios of 1.5 and 2, followed by a declining tail to larger period ratios.  Much of the tail of the distribution, beyond ratios of a few is likely contaminated by unobserved planets interspersed among observed ones.  Some of the more prominent narrow-band features are the peak near the 3:2 MMR, a modest peak near 1.7, and a cluster of peaks near the 2:1 MMR---including a large trough just interior to 2:1.  In addition, there appears to be a sizeable peak just interior to the 5:1 MMR and a slight plateau just interior to the 3:1 MMR.

We conducted a Monte Carlo simulation in order to assess the statistical significance of these features.  This test focused on period ratios between 1 and 5 and was done as follows.  First, we generated a new KDE for the period ratio distribution to serve as our ``baseline'' distribution.  This distribution used a larger bandwidth in order to smooth out the observed features (0.1 times the observed period ratio instead of 0.005).  We chose this value because it is the smallest value that yields a distribution with a single peak between period ratios of 1 and 5.

Our next step was to draw $10^4$ samples of 645 random variates from the baseline distribution (there are 645 measured, adjacent period ratios in the region of interest).  From these $\sim 6\times 10^6$ realizations we construct the resulting period ratio distribution using our nominal bandwidth of 0.005 times the period ratio.  We evaluate the resulting trial distribution at each sampled period ratio and half-way between each sampled period ratio in order to identify the values of the distribution at its associated peaks and the troughs.  From these values of local minima and maxima, we identify the curves under which 97.5\% of the relative maxima lie and over which 97.5\% of the relative minima lie.  We calculate similar curves for 99.5\% of the relative extrema.

\begin{figure}
\includegraphics[width=0.45\textwidth]{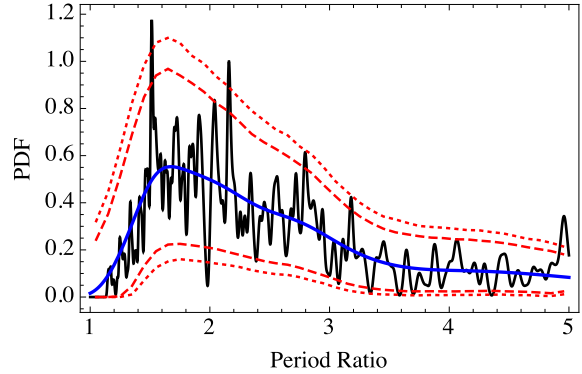}
\caption{Period ratio distribution from adjacent planet pairs between ratios of 1 and 5.  The black curve is the distribution corrected for geometric bias and pipeline completeness.  The blue, smooth curve is generated by requiring the smoothing length to produce a single relative maximum in the distribution.  This curve is the baseline distribution used in our Monte Carlo simulations to estimate the statistical significance of the different peaks and troughs.  The 95\% (dashed) and 99\% (dotted) contours from that simulation are also shown---95\% and 99\% (respectively) of the peaks and troughs from the simulation lie between those contours at each period ratio.  Analysis of the simulation, shown in Figure \ref{adjpsandts}, implies with 90\% and 99\% confidence that no more than one or two (respectively) of the three peaks that lie above or of the four troughs that dip below the dotted curves are of statistical origin.\label{mcpdist}}
\end{figure}

The nominal period ratio distribution, the baseline distribution used in the Monte Carlo simulation, and the 95\% and 99\% confidence curves are shown in Figure \ref{mcpdist}.  We claim that most peaks in the nominal period ratio distribution that reach or exceed these confidence contours are worth further theoretical and observational consideration.  We also note that our treatment likely yields a baseline distribution that over-estimates the true distribution for period ratios below $\lesssim 1.5$ where the true distribution tends toward zero.

As a means to verify our approach and the results of this simulation, we conducted a second Monte Carlo simulation where we sampled period ratios directly from the observed distribution (that is, using raw counts that are smoothed but not corrected for geometry and pipeline).  We then selected planet properties from among systems with similar period ratios (e.g., planet radius and the ratio of semi-major axis to stellar radius), then applied the geometric and pipeline corrections, and finally tabulated the minima and maxima of the resulting trial distributions.  This approach produced confidence contours similar to those shown in Figure \ref{mcpdist} indicating that effects of Poisson fluctuations in the raw planet candidate counts do not have a significant effect.

Since there are many peaks and troughs in the nominal period ratio distribution, we may expect several peaks to exceed the confidence contours (e.g., 1 in 20 should exceed the 95\%).  However, since the number of peaks in a distribution is not known \textit{a priori}, we again use our Monte Carlo simulation to estimate the number of peaks that we would expect to exceed these contours in a single realization---focusing specifically on the 99\% contours.

\begin{figure}
\includegraphics[width=0.45\textwidth]{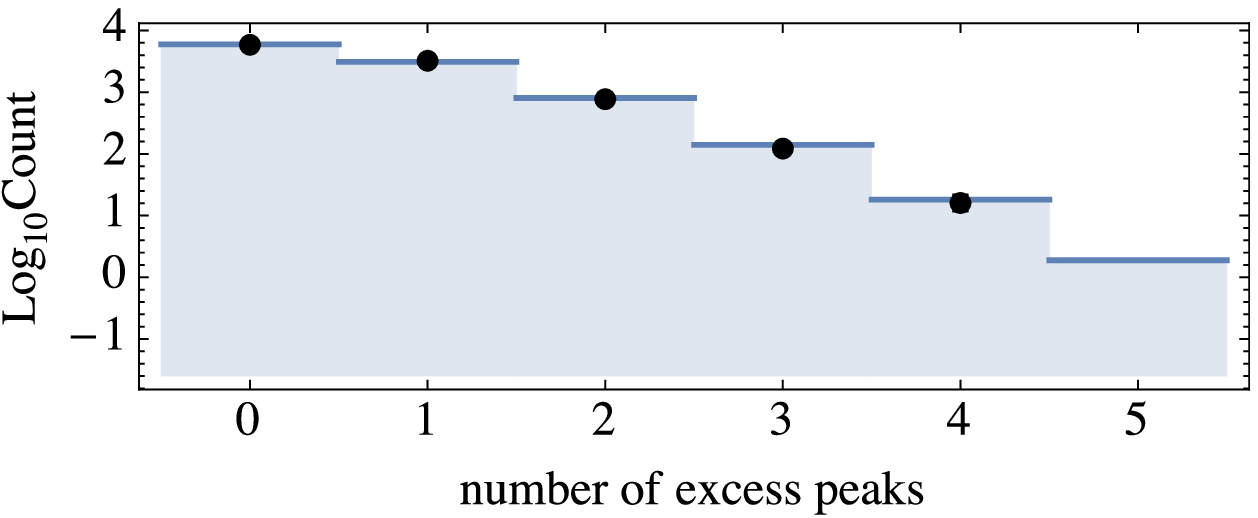}
\vskip0.1in
\includegraphics[width=0.45\textwidth]{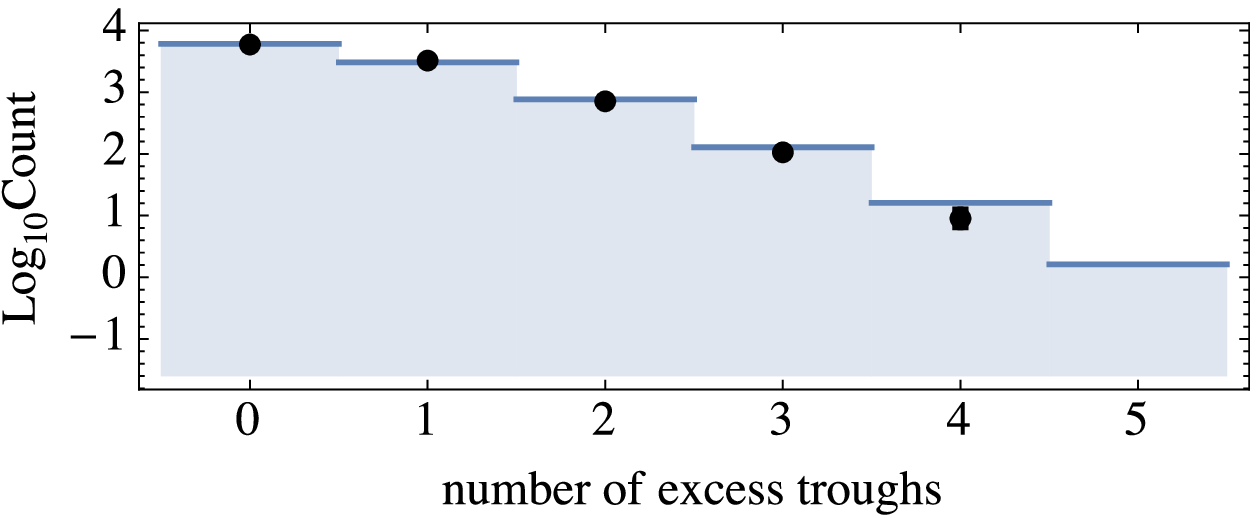}
\vskip0.1in
\includegraphics[width=0.45\textwidth]{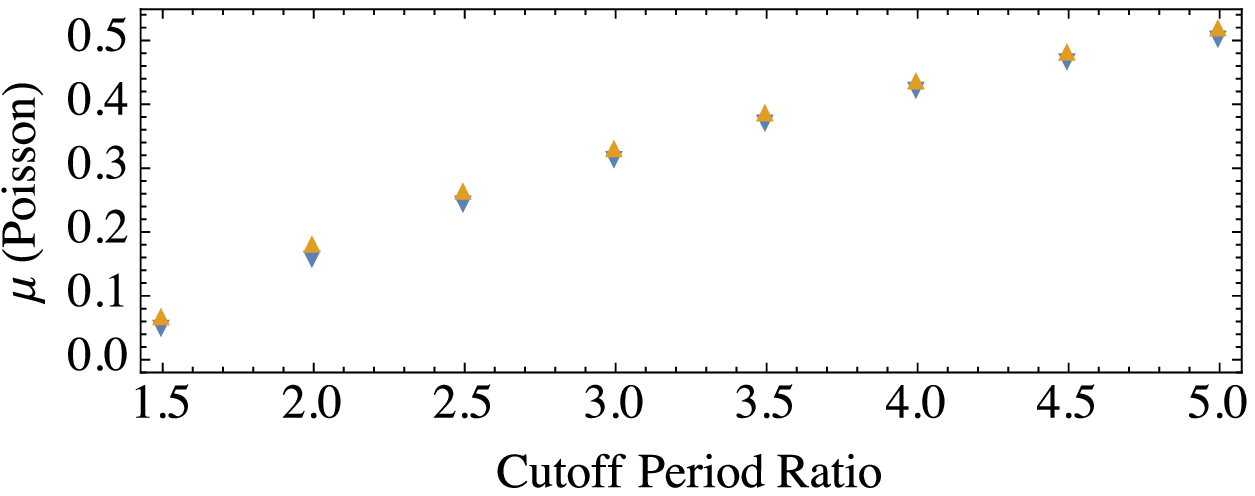}
\caption{Top and Middle: Measured number of statistical fluctuations from $10^4$ realizations of 645 planet pairs that exceed the 99\% confidence contours shown in Figure \ref{mcpdist}.  The top panel shows peaks while the middle panel shows troughs.  Estimates from the best fitting Poisson distribution are the shaded histograms with expectation values of 0.52 for the number of excess peaks and 0.50 for the number of excess troughs.  These results indicate with $\sim$90\% confidence that statistical fluctuations will produce no more than one of the three peaks or of the four troughs that exceed the dotted curves in Figure \ref{mcpdist}.  With 99\% confidence no more than two of the peaks or troughs are statistical while the most likely number of excess peaks or troughs arising from statistical fluctuations is zero.  Bottom: Expected number of peaks (upward-pointing triangles) and troughs (downward-pointing triangles) for samples of planet pairs between period ratios of 1.2 and a maximum cutoff ratio as a function of the cutoff ratio.\label{adjpsandts}}
\end{figure}

From the $10^4$ Monte Carlo realizations we count the number of times that the resulting distribution exceeds (above or below) the 99\% contour for any period ratio between 1.2 and 5.  These results are shown in Figure \ref{adjpsandts}.  We truncate our analysis to period ratios above 1.2 due to numerical artifacts that arise due to the interpolation of the confidence contours below that value.  The number of peaks and troughs is approximately Poisson distributed (also shown) with an expected value of $\simeq 0.5$ for both the maxima and the minima (0.52 and 0.50 respectively).  From these results we conclude that the most likely scenario is that none of the peaks are due to statistical fluctuations and with $\sim 90$\% and $\sim 99$\% confidence that no more than one or two (respectively) of the three peaks in the period ratio distribution that exceeds the upper dotted line in Figure \ref{mcpdist} is due to statistical fluctuations.  Similarly, for the relative minima we conclude with $\sim 90$\% and $\sim 99$\% confidence that no more than one or two (respectively) of the four troughs that dip below the line originate from statistical fluctuations while the most likely scenario is that none are statistical in nature.  We note that since the Poisson distribution is discrete, the stated confidences are approximate.

We also investigate the effects of our choice to use period ratios between 1.2 and 5 instead of a different region.  We examined the results of our Monte Carlo over a different range of values---truncating our results to several period ratios that are less than 5.  Those results are shown in the bottom panel of Figure \ref{adjpsandts}.  We see that the mean value of the Poisson distribution falls as the maximum period ratio is decreased.  The bottom panel of Figure \ref{adjpsandts} shows the mean value of the fitted Poisson distribution for period ratios greater than 1.2.  Here we see that the resulting differences also favor lower numbers of large variations.  Given these observations, our stated estimates for the number of large peaks or troughs from statistical fluctuations are conservative.  (For example, had we elected to consider only period ratios up to three, the expected number of statistical excesses drops from 0.5 to 0.3 and the probability of more than one peak being of statistical origin is less than 0.05.)

From these simulations, we identify several interesting features in the period ratio distribution---some of which were already known.  The most significant deficit is the known feature just interior to the 2:1 MMR \citep{Lissauer:2011b,Fabrycky:2014} while the most prominent peak is near the 3:2 MMR.  There is a sizeable peak just interior to a period ratio of 5.  A few modest peaks near values of 1.7, 1.9, and 2.0 are visible (as well as intervening troughs), but we cannot confidently state from our Monte Carlo that they are of physical origin---though some or all certainly may be, such as the ones near 2.0 or 1.7 \citep{Podlewska:2012,Baruteau:2013}.

Finally, and perhaps surprisingly, the prominent peak near the period ratio 2.2 is statistically significant.  When considering the likely values where one might expect peaks or troughs in the period ratio distribution, ratios like 3:2, 2:1, and even 5:1 seem plausible for dynamical reasons; 2.2 is not a value that one would have near the top of the list.  Yet, with our study, we see that the number of pairs near 2.2 is second only to the number near 3:2.  We will discuss this further in Section \ref{twopointtwo}.

\subsection{Revisiting architecture dependence on planet multiplicity}

\begin{figure}
\includegraphics[width=0.5\textwidth]{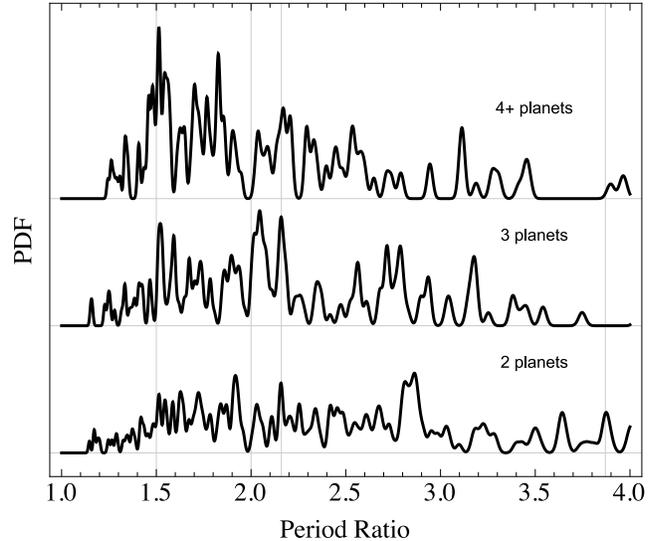}
\caption{Period ratio distribution for adjacent planet pairs in two-planet (bottom), three-planet (middle), and high-multiplicity (top) systems.  The vertical lines mark fiducial values of 1.5, 2.0, 2.16, and 3.87---locations that are either interesting dynamically, or that correspond to significant features in the period ratio distributions presented here.  \citet{Steffen:2013b} noted that the period ratio distribution of three-planet systems has a much more prominent feature near 2.0 than systems with other planet multiplicities.  This fact can also be seen here (e.g., by comparing the relative heights of the peaks near 1.5 and 2.0 for the different panels).\label{multiplicity}}
\end{figure}

A previous study of the period ratios of \kepler\ planets \citep{Steffen:2013b} shows that the distribution of period ratios of adjacent planets has some dependence on planet multiplicity---the number of planets in the system.  For example, three-planet systems tended to have a larger fraction of planet pairs near the 2:1 MMR than systems with only two planets or with four or more planets (high-multiplicity systems).  In addition, while systems with each multiplicity had a peak near 1.5, that peak was most prounounced for systems with the highest multiplicities.  These effects are seen in Figure \ref{multiplicity}, which shows the adjacent period ratio distribution for systems of differing multiplicity.

We see in this figure that, indeed, the three-planet systems have two prominent peaks between ratios of 2.0 and 2.2 while high-multiplicity systems have their largest peak near the 3:2 MMR.  The fraction of adjacent pairs with period ratios between 2.0 and 2.2 (truncating the distribution between ratios of 1 and 5) for two-planet, three-planet, and high multiplicity systems is 0.075, 0.17, and 0.12 respectively (these probabilities are obtained by integrating the (corrected) period ratio distribution between the specified values).  At the same time, the fraction of planet pairs between 1.5 and 1.65 (a similar 10\% range) is 0.073, 0.087, and 0.15---implying similar numbers of planet pairs in these windows for two-planet and high multiplicity systems, but nearly twice as many planets are near 2.0 as there are near 1.5 for three-planet systems.  (Note that there are 42 planet pairs in three-planet systems between 2.0 and 2.2.)

While systems with two planets likely have a fair number of unobserved planets with intermediate periods---making this comparison somewhat unjust---the same statement is not obviously true for systems of three planets and (for stability reasons) is even less likely for systems with four or more planets.  Thus, the discrepancy identified in \citet{Steffen:2013b} between the period ratio distributions of three-planet and high multiplicity systems remains evident and may indicate important differences in their formation or dynamical evolution.

Nevertheless, the cause of the difference between three-planet and high multiplicity systems may yet be missing, intermediate planets.  The peak near 1.5 in high multiplicity systems is quite broad and two planet pairs from this peak (provided one had a period ratio somewhat less than 1.5) would yield a period ratio for the first and third planets that lies between 2.0 and 2.2.  We show in the next section, and in section \ref{twopointtwo}, that there are several non-adjacent period ratios that land in this window.  If there were a significant number of unobserved planets in three-planet systems with period ratios near 1.5 that could explain the large number of planets observed between 2.0 and 2.2, it would mean that the peak near 1.5 is significantly larger than what is shown in Figure \ref{multiplicity} for three-planet systems.  It may also have important implications for the distribution of mutual orbital inclinations since several planets that are nominally easier to detect would have been missed.  Complementary RV observations with the specific aim to identify nontransiting, intermediate planets in these systems would be quite valuable in establishing their true nature.

\section{All planet pairs}\label{allplanets}

\begin{figure}
\includegraphics[width=0.45\textwidth]{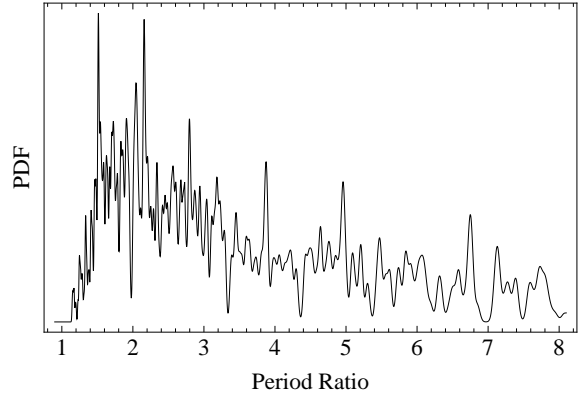}
\caption{The period ratio distribution for \kepler\ planets using all planet pairs and correcing for geometric bias and pipeline completeness.  The most prominent features are again the deficit interior to the 2:1 MMR, the peak near 3:2, and the peak near the period ratio of 2.2.\label{pdistall}}
\end{figure}

Here we investigate the distribution of period ratios when all planet pairs are considered, not just adjacent planets.  Examining all planet pairs may bring to light features in the distribution that would result from formation or dynamical processes that span multiple planets in a system.  The construction of the period ratio for all planets is identical to that for the adjacent planet pairs.  This distribution is shown in Figure \ref{pdistall}.

\begin{figure}
\includegraphics[width=0.45\textwidth]{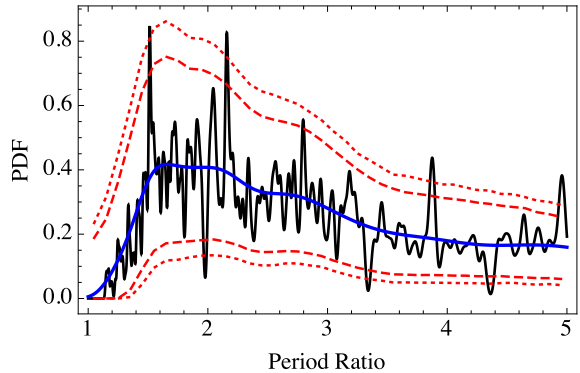}
\caption{Period ratio distribution from all planet pairs between ratios of 1 and 5.  The black curve is the distribution corrected for geometric bias and pipeline completeness.  The blue, smooth curve is generated by requiring the smoothing length to produce a single maximum in the distribution.  As with Figure \ref{mcpdist}, this curve is the baseline distribution used in our Monte Carlo simulations to estimate the statistical significance of the different peaks and troughs.  The 95\% (dashed) and 99\% (dotted) contours are shown.  Analysis of the simulation, seen in Figures \ref{adjpsandts} and \ref{cutoff}, implies with $\sim$90\% and $\sim 99\%$ confidence that no more than one or two (respectively) of the four peaks that lie above or of the three troughs below the dotted curves are of statistical origin.  The most likely number of peaks above or troughs below the dotted curve is zero.\label{mcpdistall}}
\end{figure}

We conducted a Monte Carlo analysis using a smoothed version of this distribution as was done before.  Here there are 827 planet pairs with period ratios between 1 and 5 (with $10^4$ samples we study nearly $10^7$ individual realizations).  Again the smoothing criterion is the minimum smoothing length that yields a KDE that is unimodal in the domain of interest (here it is 0.08 times the period ratio).  The results of this analysis, including the baseline distribution and 95 and 99\% confidence contours are shown in Figure \ref{mcpdistall}.

\begin{figure}
\includegraphics[width=0.45\textwidth]{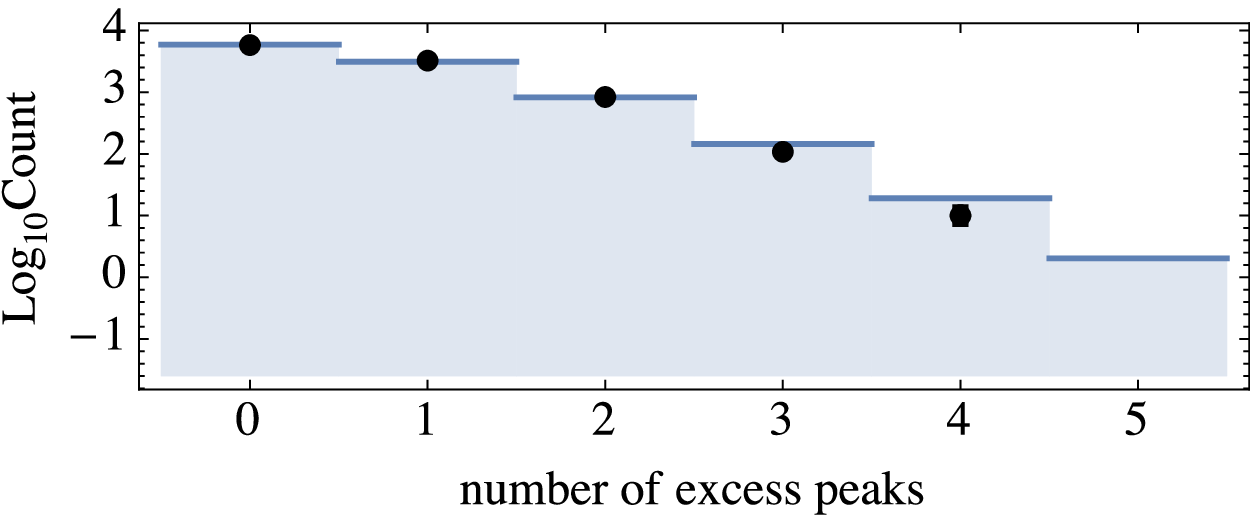}
\vskip0.1in
\includegraphics[width=0.45\textwidth]{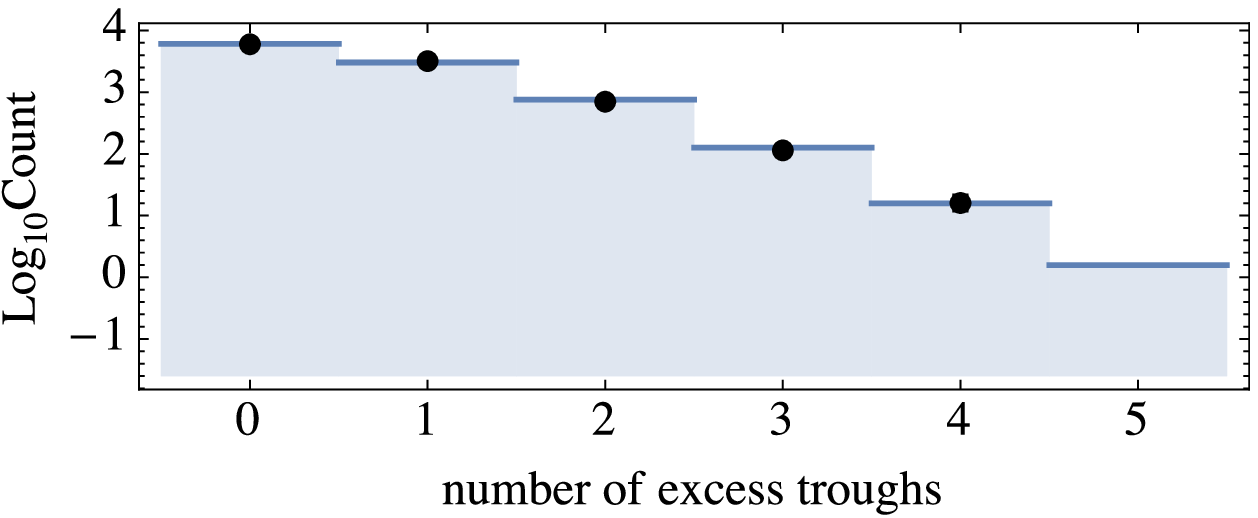}
\vskip0.1in
\includegraphics[width=0.45\textwidth]{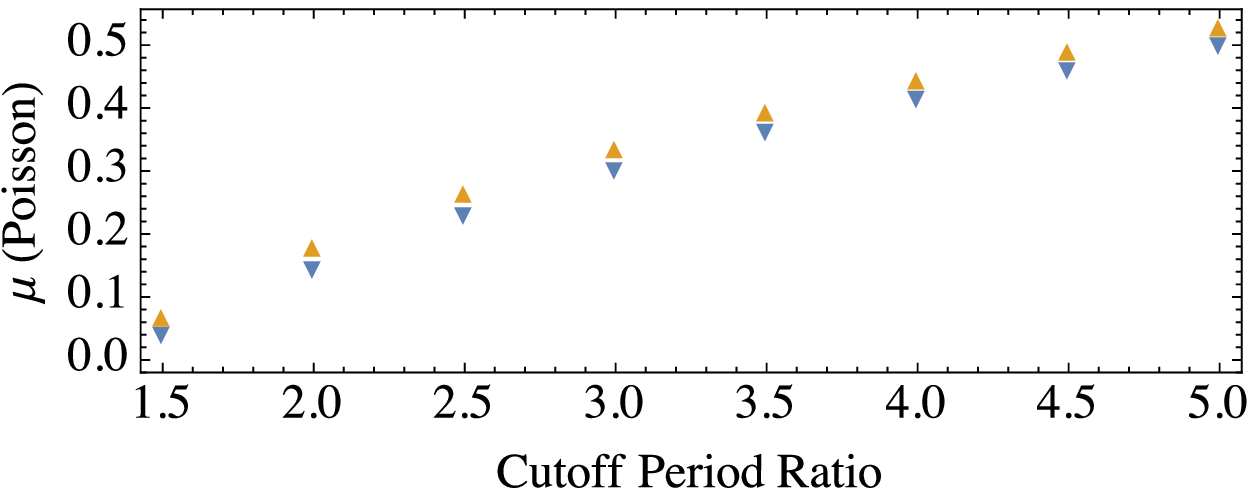}
\caption{Top and Middle: Measured number of statistical fluctuations from $10^4$ realizations of 827 planet pairs that exceed the 99\% confidence contours shown in Figure \ref{mcpdist}.  The top panel shows peaks while the middle panel shows troughs.  Estimates from the best fitting Poisson distribution are the shaded histograms with expectation values of 0.53 for the number of excess peaks and 0.50 for the number of excess troughs.  These results indicate with $\sim$90\% and $99$\% confidence that statistical fluctuations will produce no more than one or two (respectively) of the four peaks or the three troughs that exceed the dotted curves in Figure \ref{mcpdist}.  The most likely number of excess peaks or troughs from statistical fluctuations is zero.  Bottom: Expected number of peaks (upward-pointing triangles) and troughs (downward-pointing triangles) for samples of all planet pairs between period ratios of 1.2 and a maximum cutoff ratio as a function of the cutoff ratio\label{allpsandts}}
\end{figure}

The expected number of peaks in a given realization of the Monte Carlo simulation shows (again) that the number of excess peaks or troughs is approximately Poisson distributed.  Here the mean values are 0.50 for the expected number of troughs and 0.53 for the expected number of peaks.  These numbers are similar to the results for adjacent planet pairs (compare Figures \ref{adjpsandts} for adjacent pairs and \ref{allpsandts} for all pairs).

As before, the most likely number of spurious peaks or troughs is still zero while in Figure \ref{mcpdistall} we observe four peaks above, and three troughs below, the 99\% confidence contour.  The trough that was significant in the adjacent pairs that is no longer significant was near a period ratio of 3.6.  The other three troughs, near 3.4 and 4.4 and just interior to 2:1, remain.  The new peak that appears when considering all planet pairs is just interior to 4:1, while the other three peaks, at 3:2, 2.2, and near 5:1 remain statistically significant.

As before, we consider the effects of our choice to truncate our results at a period ratio of 5 by investigating cuts at smaller values.  The results of this test are shown in the bottom panel of Figure \ref{allpsandts}.  Our conclusion is the same as for the case of adjacent planets in that we believe our estimates for the probability of statistical fluctuations producing these peaks to be robust and conservative.

\subsection{Broad features of the period ratio distribution}\label{broadfeatures}

Aside from the small-scale structure in the period ratio distribution, there are important large features.  The distribution rises over small period ratios toward a broad peak between 1.5 and 2.0, then it has a long tail toward large period ratios.  Presumably, smaller period ratios occur less either for reasons of dynamical stability or because of a lack of sufficient material from which to form planets that are so close together, or both.  Considerations of formation material may also cause of the tail toward large period ratios---in the absence of large gaps in the planet-forming disk, the mass has to go somewhere, so planets can not be too far apart (barring dynamical instabilities that cause ejections \citep{Rasio:1996}).

\begin{figure}
\includegraphics[width=0.45\textwidth]{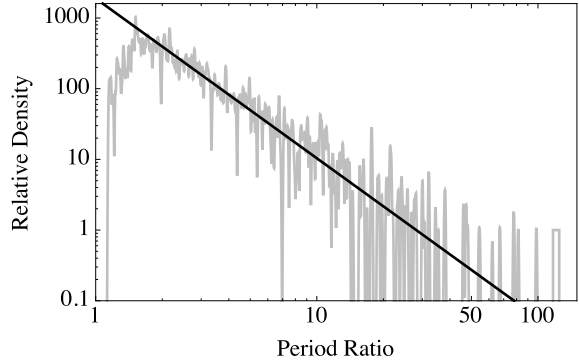}
\caption{Period ratio distribution for all planet pairs showing the power-law tail.  The fitted power law (solid line) scales with period ratio to the power $-1.26 \pm 0.05$.  We note that for period ratios $\gtrsim30$ there is typically one planet pair generating each peak.  Thus, there are large relative Poisson fluctuations in that regime.\label{pdflines}}
\end{figure}

The tail toward large period ratios follows a power-law distribution.  To estimate the exponent that describes this tail, we generate samples of 555 random variates with period ratios greater than 2.5 from the the highly smoothed baseline distribution used in the Monte Carlo simulation for all planet pairs (there are 555 planet pairs with period ratios greater than 2.5).  A maximum-likelihood fit to those data yields a PDF that is proportional to
\begin{equation}
\text{PDF}_{\prat \rightarrow \infty} \propto \prat^{-1.26 \pm 0.05}
\end{equation}
where the uncertainty in the exponent is derived by Monte Carlo with 1000 realizations of 555 variates.  This power-law estimate and the nominal period ratio distribution are shown in Figure \ref{pdflines}.

\subsection{Non-adjacent planet pairs}

\begin{figure}
\includegraphics[width=0.45\textwidth]{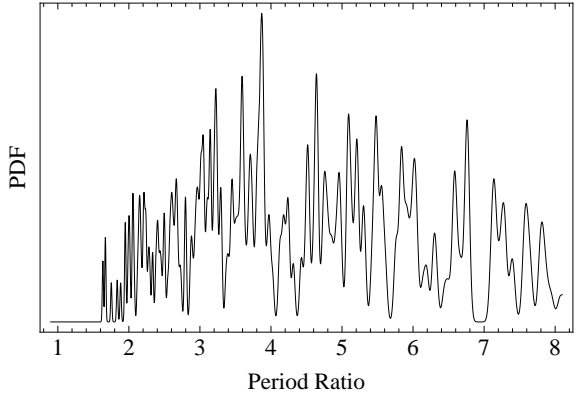}
\caption{The period ratio distribution for \kepler\ planets using only non-adjacent planet pairs and correcing for geometric bias and pipeline completeness.  The most prominent features are the spike just interior to the 4:1 MMR and the deficits exterior to the 4:1 and near the 7:1.  A Monte Carlo study of this distribution (using non-adjacent planet pairs only) shows that these features are statistically significant, but we refrain from excluding the possibility that they are statistical in origin.\label{nonadjacent}}
\end{figure}

Thus far we have examined the period ratio distribution of adjacent planet pairs and all planet pairs.  We now look at only non-adjacent plant pairs to see if there are any noteworthy features.  Figure \ref{nonadjacent} shows this distribution.  The distribution has a relatively broad peak, centered around a period ratio of 4.  The strongest, narrow-band peak is near a value of 3.9.  This peak in the non-adjacent pairs is responsible for the corresponding peak for all planet pairs in Figure \ref{pdistall}---a feature that was not very prominent in Figure \ref{pdist} using only adjacent planets only.

A Monte Carlo simulation of the non-adjacent pairs (employing a parabolic PDF over period ratios between 1 and 8) indicates that only one peak and one trough exceed the 99\% contours (the peak near 4 and the trough near 7).  The expected value for Poisson fluctuations above or below the 99\% confidence contours is quite small ($\lesssim 10^{-2}$ each).  Nominally, this analysis would indicate, with high confidence, that the peak is physical rather than statistical in nature.  However, since we only have these two, singular examples, both of which have their extrema near the significance threshold, we will not make this claim based solely on the sample of non-adjacent planets.

\section{The peaks near 2.2 and 3.9}\label{twopointtwo}

\subsection{Comments on the peak near 2.2}

Let us examine more carefully the systems that comprise the peaks near 2.2 and 3.9.  Consider a few possible explanations for the peak near 2.2.  Perhaps the peak is the result of unobserved planets on intermediate orbits.  The fact that the peak is so narrow presents a slight challenge to this explanation---unless the constituent period ratios themselves originate from a prominent narrow peak.  A pair of period ratios that results in a product of 2.2 would have typical values of $\sqrt{2.2} = 1.48$.  This value is near the peak at 3:2, however it is on the wrong side of the resonance---the peak near 3:2 is outside the 3:2.  And the peak near 2.2 is neither near nor exterior to a value of $1.5^2 = 2.25$---this fact can be seen clearly in Figure \ref{pdistshort}.  So, if the peak near 1.5 does contribute to the peak near 2.2, it would require corresponding planet pairs that are interior to that peak.  We examine this possibility shortly.

\begin{figure}
\includegraphics[width=0.45\textwidth]{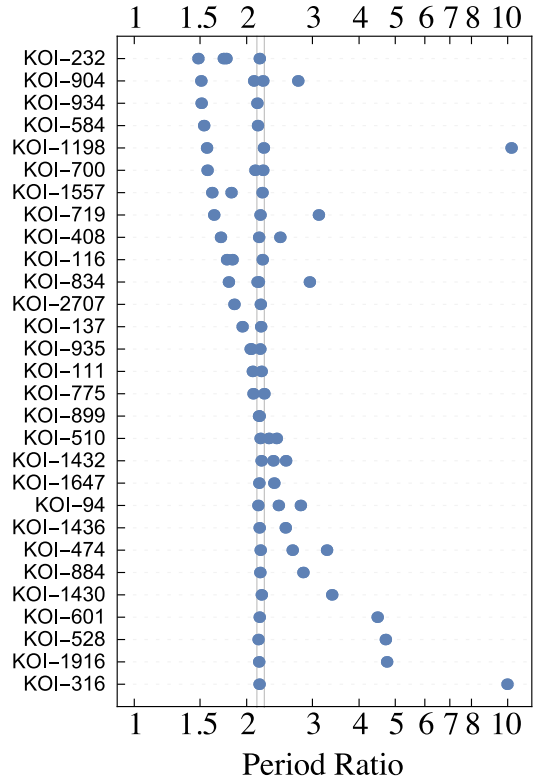}
\caption{The orbital architectures of multiplanet systems that have an adjacent planet pair in the peak near 2.2 (between 2.13 and 2.23---the two vertical lines).  Shown are the period ratios of the all adjacent planet pairs for each system, sorted in order of the minimum period ratio in the system.  There are several systems that have additional planet pairs near period ratios of 1.5, 1.7, 2.2, and between 2.5 and 3.0.  However, at this time we do not claim a statistically significant correlation between these clusters and the period ratio of 2.2 (i.e., that systems with a period ratio of 2.2 are more likely to have a second period ratio near these values).\label{multispeak}}
\end{figure}

We show in Figure \ref{multispeak} the period ratios of all adjacent planet pairs in systems where at least one pair of adjacent planets has a period ratio near 2.2 (between 2.13 and 2.23---the locations of the nearest local minima---the actual peak maximum, using our prescription, is 2.16 with the peak being more broad toward longer period ratios).  These systems are ordered by the smallest period ratio in the system.  We can see in this figure that a few of the systems have multiple planet pairs near, if not in, this narrow selection window.  There are also a small set of systems where there are a pair (or two pairs) of planets with period ratios near 1.5 or 1.7 and perhaps another small set with period ratios approaching 3:1.  However, there is no obvious second period ratio in these systems that correlates strongly with the period ratio of 2.2---if there were, we would see a large number of points at that second period ratio.

\begin{figure}
\includegraphics[width=0.45\textwidth]{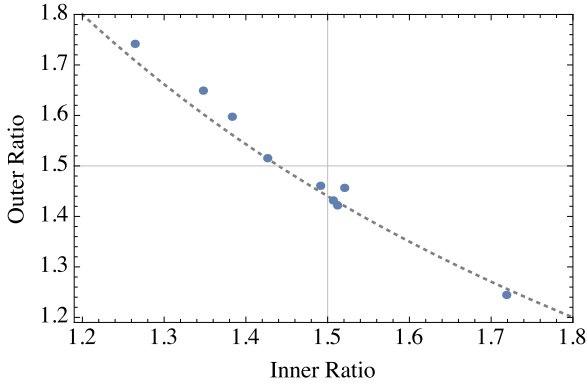}
\caption{The period ratios of planet pairs whose products are near 2.2.  All nine of these are from adjacent trios of planets (not necessarily from systems with only three planets) where the first and third planets have a period ratio near 2.2.  The location of the peak maximum is at a value of 2.16, the curve is the locus of points whose product is 2.16.  We note that the peak has a slight broadening toward larger period ratios, so the points tend to lie above (or to the right) of this curve.  The vertical and horizontal lines are for reference and show the location of the period ratio 1.5.\label{newpeaks}}
\end{figure}

When we look at all planet pairs, not just adjacent ones, the peak near 2.2 gets larger---roughly nine new planet pairs contribute.  So, indeed there are non-adjacent planets that add meaningfully to the height of the peak at 2.2.  The three-planet chains whose period ratios multiply to 2.2 are shown in Figure \ref{newpeaks}.  There is a small cluster where the two adjacent ratios are near 1.5 (one pair is near 1.5 while the other is slightly interior to 1.5).  Four of the combinations do not have a planet pair that is particularly close to 1.5.  Thus, while intermediate planets that form period ratios near 1.5 do make a contribution to the peak at 2.2, they do not account for all pairs that form the peak.  Nevertheless, the fact that 5 of the 9 systems have at least one planet pair with a period ratio in the range $1.5 \pm 0.02$ may indicate that the peak originates from the combined contributions of an overall smooth distribution and planet pairs from the peak near 1.5.

If the peak near 2.2 has a physical or dynamical origin, it is not obvious what that origin is.  One possible explanation may be seen in \citet{Baruteau:2013} (see also \citet{Podlewska:2012}).  There, the authors attempt to explain the period ratio of 1.7 measured in Kepler-46 \citep{Nesvorny:2012} as a byproduct of the interaction of the two planets with a gaseous disk.  While the masses of the planets in those simulations are likely a bit larger than the typical \kepler\ planet, a number of their simulations produced a period ratio near 2.2 instead of the targeted 1.7.  Those simulations that stopped near 2.2 retained an annulus of residual gas between the two planets that enabled repulsive interactions between the planets and the wakes they produce in the disk.  The simulations that bypassed 2.2 were able to deplete that region of its gas and continue their migration to smaller period ratios.  If this mechanism is the cause of the observed, narrow feature at 2.2 it would imply either a relatively uniform outcome from a variety of initial conditions or a relatively uniform set of initial conditions.

\subsection{Speculations on the peak near 3.9}

Another statistically significant peak shown in Figure \ref{pdistall} that lies just interior to the 4:1 MMR (at a period ratio of 3.87).  This peak is due in large part to non-adjacent planet pairs (as can be seen in Figure \ref{nonadjacent}).  In all there are 19 planet pairs that contribute to this peak (the ``peak'' being period ratios between 3.78 and 3.93---the locations of the nearest local minima from our KDE of the period ratio distribution).  Of these planet pairs, six are observed as adjacent pairs with five appearing in two-planet systems (KOIs 119, 291, 1435, 2168, and 2554) and one appearing in a five-planet system (KOI 952) which has both an adjacent pair and a non-adjacent pair.  The remaining 12 non-adjacent planet pairs with this period ratio appear in 11 different multiplanet systems (KOIs 82, 116, 152, 250, 671, 707, 834, 898, 1336, 1426, and 1895; with KOI 834 having two separate non-adjacent planet pairs that contribute).  We note that because the actual location of the peak maximum is at a period ratio of 3.87, the constituent systems are quite unlikely to be dynamically associated with the 4:1 MMR today.

\begin{figure}
\includegraphics[width=0.45\textwidth]{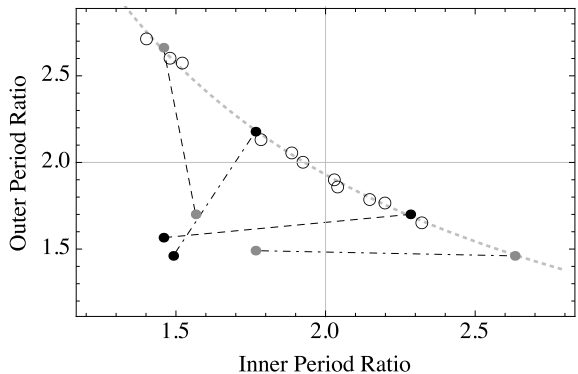}
\caption{The period ratios of planet pairs whose products are near 3.9 (similar to Figure \ref{newpeaks}).  Open circles mark the period ratios of adjacent trios of planets where the first and third planets have a period ratio near 3.9.  The curve is the locus of points whose product is 3.87 (the value at the peak maximum).  Most planet pairs that contribute to the peak have period ratios near 2.  A few pairs have a small period ratio near 1.5 and a larger one near 2.5.  The black dots correspond to sequences of four planets, lines connect the smallest two period ratios (in the lower left) to the combined product of all three ratios.  That is, the smallest two ratios are combined into one, then that product is treated as either an inner ratio or outer ratio (as appropriate) and combined with the largest ratio to produce the point near the curve.  Gray dots correspond to the same sequences of four planets, but the largest and its neighboring ratio are combined first, then that product is used to produce the point near the curve.  Dashed lines correspond to KOI 82 while dot-dashed lines correspond to KOI 671.  The vertical and horizontal lines are for reference and show the location of the period ratio 2.0.  The combination of period ratios from KOI 82 that produces the gray point among the island of points in the upper left may suggest that those systems harbor an additional, undetected planet (one that yields an architecture similar to KOI 82).
\label{peak41}}
\end{figure}

Figure \ref{peak41} shows the period ratios for non-adjacent planet pairs whose product is $\simeq 3.87$.  Also shown are different possibilities for interpreting sequences of three period ratios that have the same product (these come from KOIs 82 and 671).  The clustering of observed two-ratio combinations that make a period ratio of 3.87 hint at a slight concentration of both period ratios around $\sim$2.  A small island of points exists for combinations near 1.5 for the inner pair and 2.5 for the outer.

If we examine the three-ratio combinations that produce 3.87 (there are two of them), they both come in sequences with a wide ratio near 1.7 and two adjacent smaller ratios, each near 1.5.  If we construct an ordered pair by combining the two adjacent, smaller ratios into a single ratio, then the resulting points in Figure \ref{peak41} are more consistent with the bulk of the two-ratio systems.  If, on the other hand, we construct the ordered pair by combining the largest ratio with its neighboring smaller ratio, the resulting points are farther from the bulk of the two-ratio systems.  Interestingly, combining the large ratio with a small ratio in KOI 82 does produce a point that lands in the island of points mentioned in the previous paragraph.

These observations lead to a few ideas that might be worth further consideration.  First, the observed peak near the period ratio of 3.9 may be real (non-statistical in origin) and may result from combinations of ratios from other peaks in the overall distribution (e.g., the peak near 1.5, which contributes the third ratio in both the KOI 82 and the KOI 671 systems).  Second, the three systems with points that are isolated in the small island of Figure \ref{peak41}---KOIs 250, 952, and 1336---may contain unobserved, intermediate planets.  Given the scenario of KOI 82, we speculate that the intermediate planet would have a period ratio near 1.5 with respect to one of the two known planets.  Third, the systems that contain adjacent planet pairs with period ratios near 3.9 may also have unobserved, intermediate planets interleaved among the known planets.

\section{First-order MMRs}\label{firstorders}

Previous studies of the architectures of the \kepler\ systems have shown a significant asymmetry near first-order MMR.  Specifically, there is a lack of planet pairs just interior to the resonance and an excess exterior to the resonance.  There are a variety of ways to measure ``nearness'' to a resonance and the utility of each depends upon the question one wishes to address.  Three have recently appeared in the literature, which we will call $\Delta, \ \epsilon$, and $\zeta$.  (Note that $\Delta$ here does not correspond to the number of mutual Hill radii between planets---which often uses this same symbol.)

These three quantities are used, for example, in \citet{Lithwick:2012a}, \citet{Delisle:2014}, and \citet{Fabrycky:2014} respectively and are shown in Equation~\ref{quantities},
\begin{equation}\label{quantities}
\begin{split}
\Delta &= \prat \frac{j}{j+1} -1 \\
\epsilon &= \prat - \frac{j+1}{j} \\
\zeta &=  2\left( \frac{j}{\prat -1}-\text{Round}\left[ \frac{j}{\prat -1}\right] \right),\\
\end{split}
\end{equation}
where $\prat$ is the ratio of orbital periods (greater than unity).  $\Delta$ is used in \citet{Lithwick:2012a} to calculate the deviations from a constant orbital period that arise from planet-planet interactions (Transit Timing Variations or TTVs \citep{Agol:2005,Holman:2005}).  The quantity $\epsilon$ is used in \citet{Delisle:2014} and \citet{Chatterjee:2014} to measure the distance from MMR.  The latter paper uses $\epsilon$ to measure the displacement of a planet from MMR that arises from interactions with a disk of planetesimals as a means to explain the observed asymmetry near MMR.

\begin{figure*}
\includegraphics[width=0.9\textwidth]{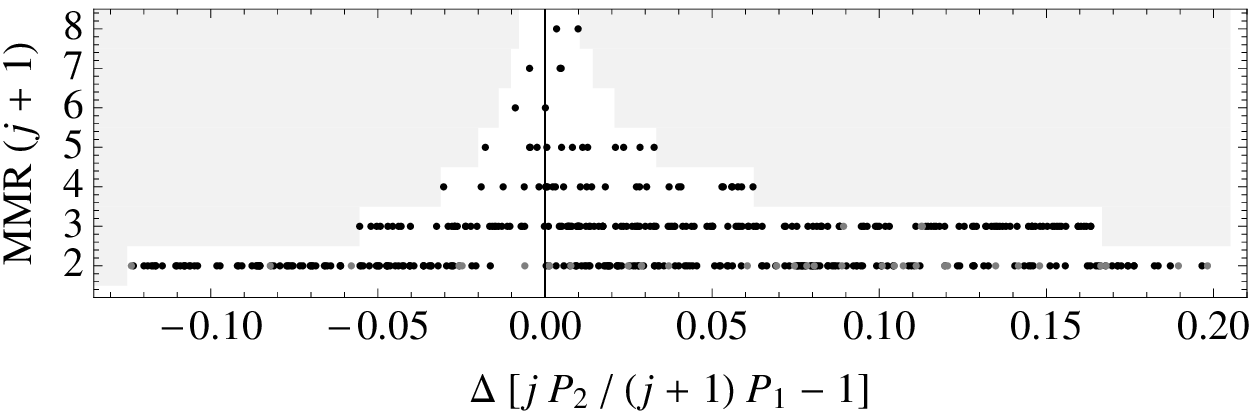}

\bigskip
\includegraphics[width=0.9\textwidth]{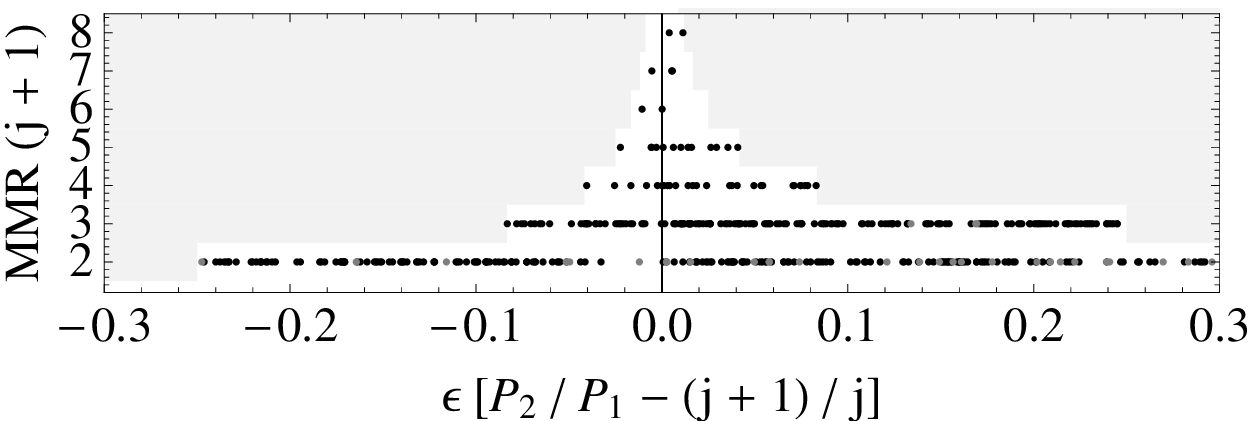}

\bigskip
\includegraphics[width=0.9\textwidth]{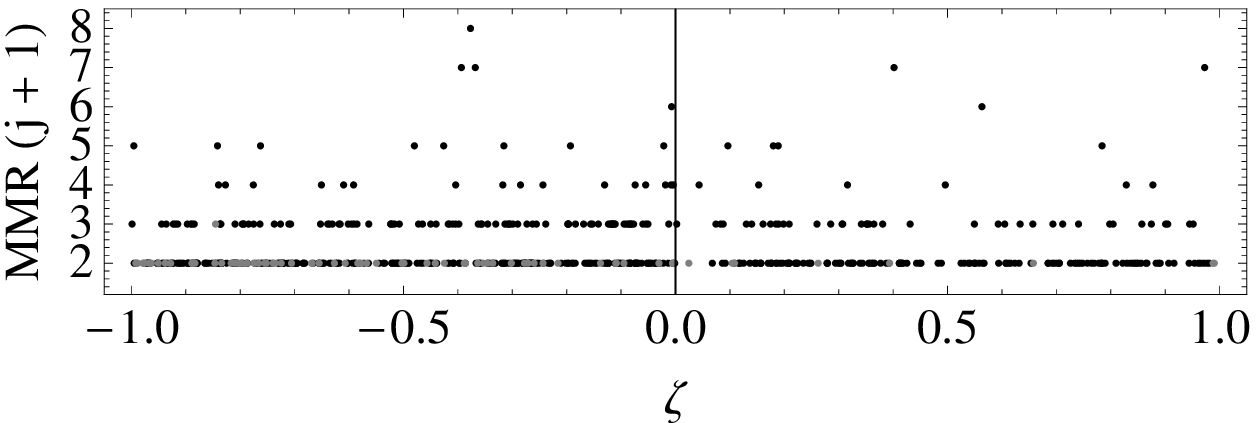}
\caption{The distance from first-order MMR for different measurement quantities.  Adjacent planets are shown with black dots while non-adjacent planets are gray dots.  The gray shaded regions indicate the boundaries for each MMR where each quantity is forbidden.  That is, a planet pair in the shaded region for one MMR is closer to a neighboring MMR and appears there.  The top panel shows the quantity $\Delta$ used, for example, in \citet{Lithwick:2012a,Hadden:2014}.  The middle panel shows the quantity $\epsilon$ used in \citet{Delisle:2014,Chatterjee:2014}.  $\Delta$ and $\epsilon$ differ by a factor of $(j+1)/j$.  The bottom panel shows the quantity $\zeta$ used in \citet{Lissauer:2011b, Fabrycky:2014}.  Each of these quantities may be useful for different purposes.  We note that the deficit of planet pairs interior to first-order MMR is more prounounced using $\zeta$ and is nearly the same size for all indices $j$.  Notable planet pairs that lie in that gap are two non-adjacent pairs from KOI-730 near the 2:1, a pair from KOI-1599 near the 3:2, and a pair from KOI-430 near the 4:3.\label{zetapic}}
\end{figure*}

The quantity $\zeta$ was derived in \citet{Lissauer:2011b} to provide a means to combine, or ``stack'', the distribution of period ratios for all first-order MMRs in a meaningful way.  Its generalization, explained in more detail in \citet{Fabrycky:2014}, applies to MMRs of any order.  $\zeta$ stretches the neighborhood of a given first-order MMR to span the interval $(-1,\ 1)$ where the ``neighborhood'' of one MMR runs between the nearest MMRs of the next higher order.  (For example, the neighborhood of the 3:2 MMR runs between the 5:3 and 7:5 MMRs.)  Figure \ref{zetapic} shows how all planets in the neighborhoods of first-order MMRs are distributed in terms of each of these three quantities.  The asymmetry near first-order MMR is most pronounced when using $\zeta$, but the dynamics of the system are typically described using $\Delta$ or $\epsilon$.

Another, more physically motivated, way to measure the distance from MMR for a planet pair is to consider the libration width of those planets if they were located at their nearest MMR.  To investigate this approach we will use as the width the supremum (or least-upper-bound) on the total orbital period libration width using Equation 8.76 of \citet{Murray:1999}.  This formula applies to a massive exterior planet and a test-particle inner planet.  While this approximation is not technically valid for the case of two interacting planets, the libration width generally scales with the total planet mass \citep{Deck:2013} and we expect this approximation to yield adequate results---especially since we are using a mass-radius relationship to estimate the planet masses with correspondingly large uncertainties due to density differences among the \kepler\ planets.

Our mass-radius relationship for estimating planet masses is $m = r^3 M_\oplus$ for planets smaller than 1$R_\oplus$ (yielding planets with constant, Earth-like densities) and $m=r^2 M_\oplus$ for planets larger than 1$R_\oplus$ \citep{Lissauer:2011b}.  For eccentricity, we choose eccentricities that give us the supremum or smallest-allowed maximum libration width.  Dividing the observed distance from MMR by this resonance width gives approximately the maximum value for the true distance from MMR---planets should be closer to resonance than the values specified.

\begin{figure}
\includegraphics[width=0.45\textwidth]{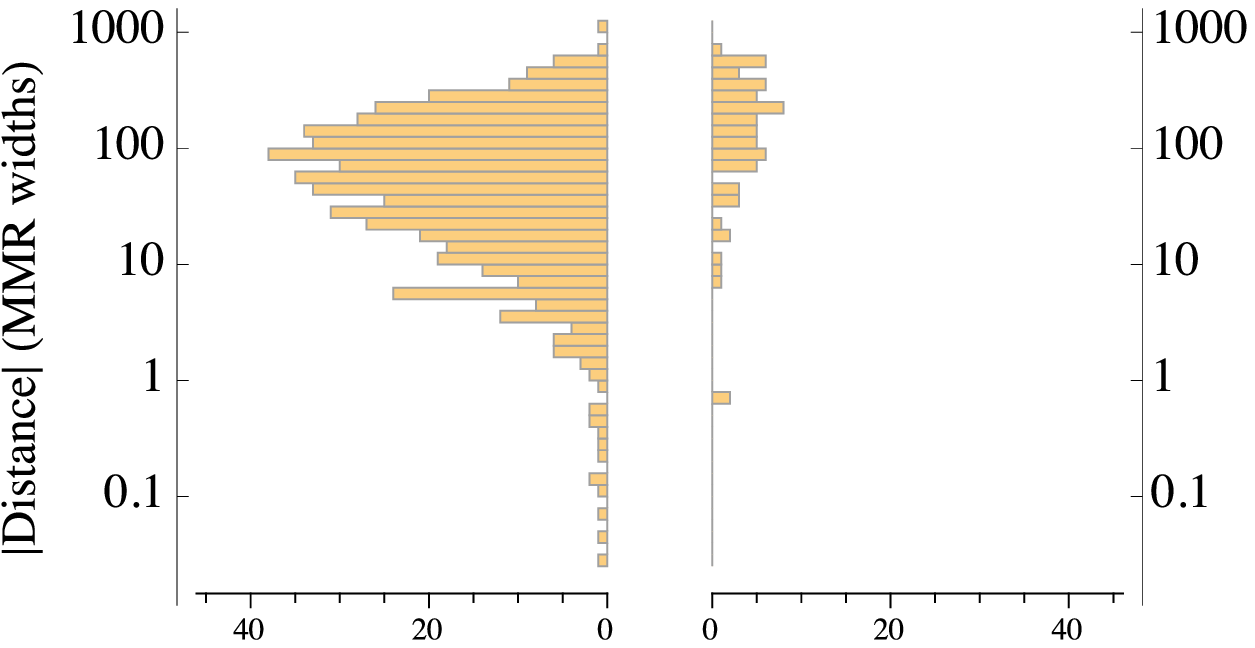}
\caption{Histogram of the absolute distance from first-order MMR for adjacent (left) and non-adjacent (right) planet pairs.  These distances are measured in units of the estimated supremum (least upper bound) of the resonance libration width (i.e., the observed distance from MMR in units of the minimum expected distance that would arise from resonant interactions).  Systems most likely to be in MMR are those with estimated distances less than a few.  Two non-adjacent planet pairs from KOI-730 are notable in the right panel and are quite close to the 2:1 MMR (though any resonant dynamics of this system would likely involve all four planets).\label{reswidthhist}}
\end{figure}

\begin{figure*}
\includegraphics[width=0.9\textwidth]{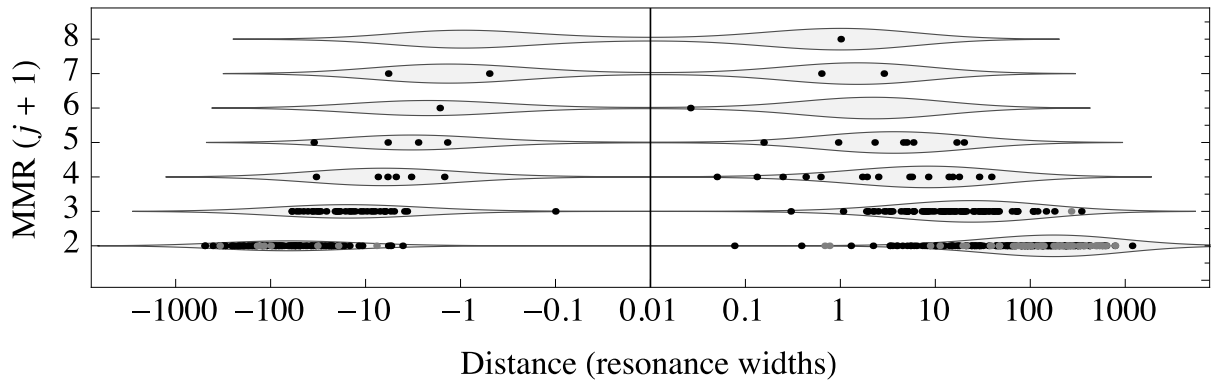}
\caption{Distance from first-order MMR when measured in units of the estimated supremum of the resonance libration width (i.e., the observed distance from MMR in units of the minimum expected distance that would arise from resonant interactions).  Black dots indicate adjacent planet pairs while gray dots indicate non-adjacent planet pairs.  The gray shaded regions are the distributions that one gets if planet pairs are logarithmically distributed in the neighborhoods of each MMR (planet sizes, from which masses are derived, are distributed log-normally from the best fitting parameters to the \kepler\ data).  We note that the tails closer to MMR tend to be more populated than the tails farther from MMR.\label{reswidthpic}}
\end{figure*}

The results of these measurements are shown in Figures \ref{reswidthhist} and \ref{reswidthpic}.  Figure \ref{reswidthhist} is a histogram of the absolute value of the distances from the nearest MMR.  This figure includes separate histograms for adjacent pairs and non-adjacent pairs.  Among the non-adjacent pairs there are only two that are particularly close to a first-order MMR.  These are the planets in KOI-730, a system that is in or near a 8:6:4:3 (4:3, 3:2, 4:3) resonance chain, the first and third, and second and fourth planets both being close to the 2:1 MMR.

Figure \ref{reswidthpic} shows the signed distances from MMR for both adjacent and non-adjacent planet pairs---now grouped by resonance index $j$.  By way of comparison, for each index we show the distribution of measured values one would get for a total of $10^4$ systems that are logarithmically spaced in period ratio and with planet radii drawn from a log-normal distribution.  The parameters for the planet size distribution come from a fit to the observed distribution with period ratios from 1 to 3---the masses are then estimated using our mass radius relationship.

From Figure \ref{reswidthpic} we can see that most planet pairs are roughly consistent with what one would expect from logarithmically spaced planet pairs in the respective neighborhoods.  We also see the deficit of planet pairs just interior to MMR, though there is one notable exception---KOI-1599 has a period ratio of 1.4997.  For pairs that are wide of MMR, the close-proximity tail is more populated than the tail that is farther from MMR, perhaps indicating a small population of planets that are indeed resonating and are therefore distributed differently.  Our claim is that planet pairs with absolute distances that are less than a few are where resonant configurations are more likely to be found (we do not claim that any specific pair is resonating).  Candidate systems less than 3 widths are given in Table \ref{closetable}.

\begin{table}
\caption{Planet pairs within three libration widths of first-order MMR.}\label{closetable}
\begin{tabular}{llllll} \hline
KOI & KOI & Kepler & MMR & Period & Distance \\ 
	& Pair$^a$	& number	&	& Ratio	& from MMR$^b$ \\ \hline \hline
K00262 	&	 01/02 	&	50	&	 6:5 	&	1.20014	&	 0.0263 \\
K00730 	&	 01/03 	&	223	&	 4:3 	&	1.33357	&	 0.0499 \\
K01426 	&	 02/03 	&	297	&	 2:1 	&	2.00239	&	 0.0764 \\
K01599 	&	 02/01 	&	 -- 	&	 3:2 	&	1.49971	&	 -0.1003 \\
K00730 	&	 04/02 	&	223	&	 4:3 	&	1.33379	&	 0.1317 \\
K02086 	&	 01/02 	&	60	&	 5:4 	&	1.25068	&	 0.1553 \\
K02086 	&	 02/03 	&	60	&	 4:3 	&	1.33436	&	 0.2465 \\
K00730 	&	 02/01 	&	223	&	 3:2 	&	1.50156	&	 0.2996 \\
K01338 	&	 03/02 	&	 -- 	&	 2:1 	&	2.0007	&	 0.3873 \\
K00523 	&	 01/02 	&	177	&	 4:3 	&	1.34073	&	 0.4313 \\
K01070 	&	 03/02 	&	266	&	 7:6 	&	1.16127	&	 -0.4982 \\
K00314 	&	 03/01 	&	138	&	 4:3 	&	1.33641	&	 0.6179 \\
K00277 	&	 02/01 	&	36	&	 7:6 	&	1.17194	&	 0.6307 \\
K00730$^c$ 	&	 04/01 	&	223	&	 2:1 	&	2.00276	&	 0.6820 \\
K00730$^c$ 	&	 02/03 	&	223	&	 2:1 	&	2.00243	&	 0.7661 \\
K01576 	&	 01/02 	&	307	&	 5:4 	&	1.25618	&	 0.9477 \\
K02160 	&	 02/01 	&	 -- 	&	 8:7 	&	1.14681	&	 1.0078 \\
K00620 	&	 03/02 	&	51	&	 3:2 	&	1.52595	&	 1.0674 \\
K00377 	&	 01/02 	&	9	&	 2:1 	&	2.01864	&	 1.2882 \\
K02768 	&	 01/02 	&	404	&	 5:4 	&	1.24702	&	 -1.3814 \\
K00430 	&	 02/01 	&	 -- 	&	 4:3 	&	1.32505	&	 -1.4845 \\
K00654 	&	 01/02 	&	200	&	 6:5 	&	1.18934	&	 -1.6555 \\
K00505 	&	 02/04 	&	169	&	 4:3 	&	1.34744	&	 1.6977 \\
K02038 	&	 01/02 	&	85	&	 3:2 	&	1.5064	&	 1.8717 \\
K01858 	&	 02/01 	&	 -- 	&	 4:3 	&	1.35195	&	 1.8959 \\
K01563 	&	 01/02 	&	305	&	 3:2 	&	1.51097	&	 1.9400 \\
K00168 	&	 03/01 	&	23	&	 3:2 	&	1.51155	&	 1.9609 \\
K00934 	&	 02/03 	&	254	&	 3:2 	&	1.51034	&	 2.1975 \\
K00806 	&	 03/02 	&	30	&	 2:1 	&	2.06835	&	 2.2037 \\
K00904 	&	 02/03 	&	55	&	 3:2 	&	1.50761	&	 2.2582 \\
K00157 	&	 06/01 	&	11	&	 5:4 	&	1.26406	&	 2.2939 \\
K00248 	&	 01/02 	&	49	&	 3:2 	&	1.51489	&	 2.4925 \\
K00500 	&	 01/02 	&	80	&	 4:3 	&	1.34995	&	 2.5049 \\
K00351 	&	 06/05 	&	90	&	 5:4 	&	1.24424	&	 -2.7964 \\
K01665 	&	 02/01 	&	 -- 	&	 7:6 	&	1.17232	&	 2.8688 \\
K00886 	&	 01/02 	&	54	&	 3:2 	&	1.5069	&	 2.8766 \\ \hline\end{tabular}
$^a$ KOI numbers are not always in order of orbital period.\\
$^b$ Distances are measured in units of the minimum expected variation in period ratio from resonant interactions.\\
$^c$ Planet pairs that are not adjacent to each other.
\end{table}

\section{Period ratios near large integers}\label{integerratios}

There are several systems that have planet pairs with period ratios very close to large integers ($\geq 3$).  Some of these systems were discussed in \citet{Lissauer:2011b} which we revisit here.  Table \ref{integermmr} shows the systems that have a period ratio within 0.01 of a $j$:1 MMR.  Also shown are the number of planet pairs with period ratios in the range $j \pm 1/2$ and the number of known planets with orbital periods between the pair in question.

\begin{table}
\caption{Systems with planet pairs near $j$:1 MMR.}\label{integermmr}
\begin{center}
\begin{tabular}{rrrrc} \hline
KOI & MMR & $\prat - j$ & \# near  & \# obs. \\ 
      & $j$:1 &	                & $j \pm 1/2$     &  between \\ \hline \hline
K00117 &  3   &    0.0091 &  195      & 1 \\
K01835 &  3   &    0.0098 &  195      & 1 \\
K01101 &  4   &  -0.0003 &    91      &    \\
K00657 &  4   &    0.0012 &    91      &    \\
K02715 &  5   &  -0.0019 &    68      & 1 \\
K01574 &  5   &    0.0024 &    68      &    \\
K00812 &  6   &    0.0058 &    35      & 1 \\
K00408 &  9   &  -0.0076 &    18      & 2 \\
K01336 &  9   &  -0.0052 &    18      & 2 \\
K00070 &  21 &  -0.0019 &      6      & 3 \\ \hline
\end{tabular}
\end{center}
\end{table}

From the counts in Table \ref{integermmr} one may estimate the expected number of planet pairs with period ratios as close or closer to the corresponding integer assuming uniformly distributed planet pairs (following the prescription of \citet{Lissauer:2011b}).  Doing so would indicate that in many cases, the proximity to MMR is quite small and consequently that there is some significance to the observed proximity to MMR.  However, we caution, and show below, that such conclusions can not be drawn with these observations---at least in this manner.

One interesting fact about many of these systems is that there are planets with intermediate orbital periods, but these orbital periods generally are not near any MMR.  Consider KOI-812.  \citet{Lissauer:2011b} speculated that the 6:1 MMR for this system is relatively weak without a sizeable orbital eccentricity, but that an intermediate planet with a period that formed a 2:1/3:1 chain would be much stronger and a 2:1/2:1/3:2 combination even more so.  An intermediate planet was indeed discovered in the Q8 data, but the period ratios formed with the new planet are 2.34 and 2.56---near nothing of particular consequence.

Perhaps a better example of a system with intermediate planets is KOI-70 (Kepler-20, \citet{Gautier:2012,Fressin:2012}).  The period ratio of the innermost known planet and the outermost known planet is 20.9981.  And, assuming uniformly distributed planet pairs between period ratios of 20.5 and 21.5, we would expect to find 0.02 pairs this close to the 21:1 MMR.  There are three intermediate planets in the system.  Yet, as with KOI-812, the period ratios in this system of planets do not imply strong dynamical interactions: 1.65, 1.78, 1.80, and 3.96.  The same is true for KOIs 408 and 1336, both have two outer planets with period ratios of 8.9924 and 8.9948 respectively.  Each system has two intermediate planets yielding period ratios of 2.15, 1.70, and 2.45 for KOI-408 and 2.29, 1.52, and 2.57 for KOI 1336---only one of which is close to a low-order MMR.

The fact that many of the planet pairs with period ratios very close to a $j$:1 MMR have intermediate planets with period ratios that do not favor strong dynamical interactions (such as multibody resonances or chains of resonances) cautions against over interpreting these period ratios and hint that MMR may not be a significant consideration in most multiplanet systems---even in ones where observed period ratios near large integers might motivate such a hypotheses.

We test the hypothesis that $j$:1 MMRs are preferred by counting the number of planet pairs with period ratios near random offsets from the integers.  To do this, we ran 50 simulations.  For each simulation we chose a random number between $\pm 1/2$ and added it to the integers that are greater than or equal to three (though our conclusions are similar if we consider larger lower bounds).  We count the number of period ratios from all planet pairs that are within $\pm 0.01$ of each set of locations.  Thus, if one of the random numbers was 0.2, we would count period ratios the regions $3.2 \pm 0.01, \ 4.2 \pm 0.01, \ 5.2 \pm 0.01$, etc.  (Note that the size of this window means that sampling more than 50 random numbers is not justified.)

We find that the mean number of systems that lie within these small regions of interest is 11.2 with a standard deviation of 4.3.  The average of the nearest integer (i.e., mean $j$) for the observed systems is 7.4.  These numbers are entirely consistent with what is observed near the $j$:1 MMRs shown in Table \ref{integermmr}, which has 10 planet pairs near integers with a mean $j$ value of 6.9.  At very least, this test demonstrates that there is no preference for planet pairs to be near $j$:1 MMRs beyond the 2:1.  This conclusion does not imply that resonance dynamics plays no role in these systems---only that, in general, such dynamics do not play a role that is large enough to produce a noticeable feature (whether excess or deficit) in the distribution of period ratios.

\section{Sanity check with Quarter 16 data}\label{quarter16}

\begin{figure}
\includegraphics[width=0.45\textwidth]{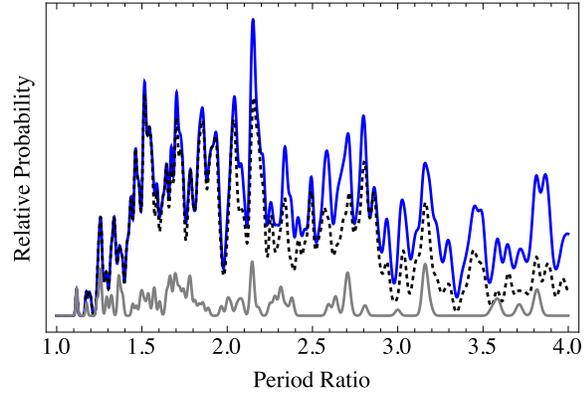}
\caption{Period ratio distribution for adjacent planet pairs (black dotted), all planet pairs (blue solid), and all planet pairs from new systems only (gray solid) using preliminary \kepler\ Q16 data.  The peak near 2.2 grows somewhat larger and the peak near 1.7 now exceeds the peak near the 2:1 MMR.  The peak near 3.9 also remains.  These results support the previous claims that the most significant of the peaks are unlikely to be of statistical origin.\label{q16pdist}}
\end{figure}

While our results focus on the Q8 catalog of \citet{Burke:2014}, which has undergone more, and more uniform, vetting than the available data from later observations, we reproduce some of our figures using data through Quarter 16 (Q16) in order to verify that the addition of new candidates does not obviously negate any of our points or conclusions above\footnote{Our Q16 data was retrieved from the Exoplanet Archive on Decemberl 10, 2014.}.  The short answer is no, our conclusions are unchanged.  Figure \ref{q16pdist} shows the period ratio distribution for adjacent and all planet pairs, an additional line shows new planet pairs.  Now, the peak near 2.2 is stronger than the peak near 1.5.  (Indeed, if one considers only new systems, not new planets in known systems, then the most prominent peak is the one near 2.2.)  The peak near 3.9 remains large.  In addition, the peak near 1.7---which we are not able to claim as statistically significant from our earlier analysis---also grows in significance and even eclipses the peak near the 2:1 MMR (possibly indicating a physical origin as well).

\begin{figure*}
\includegraphics[width=0.9\textwidth]{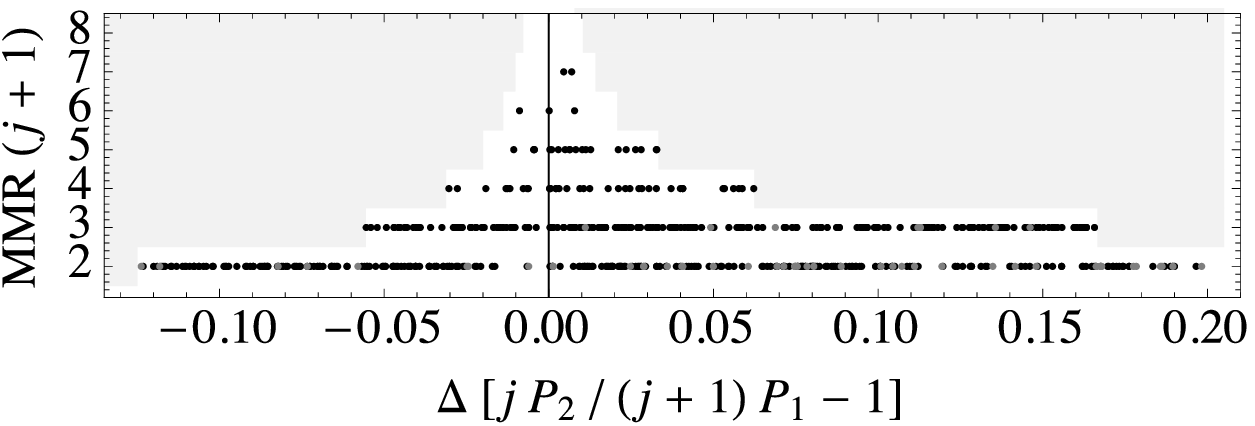}

\bigskip
\includegraphics[width=0.9\textwidth]{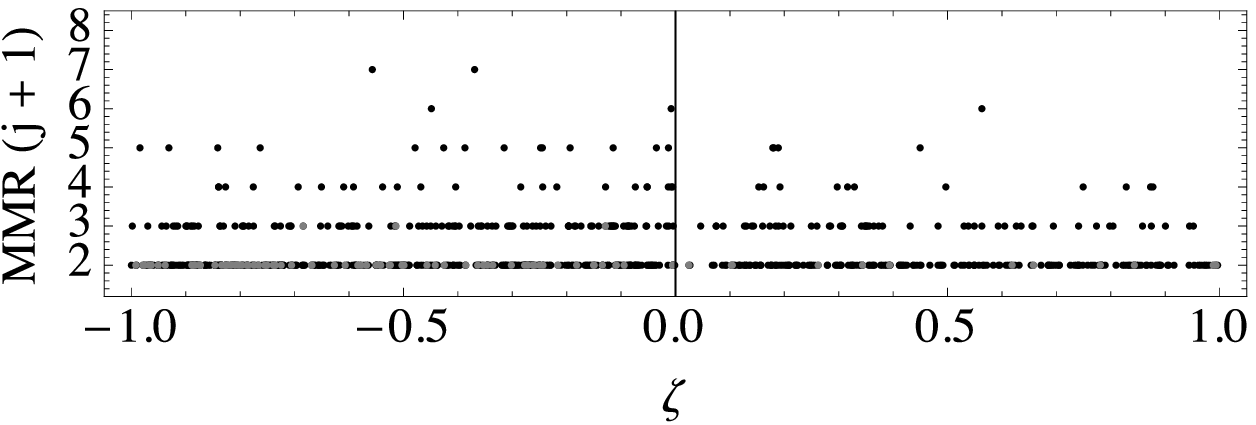}
\caption{The distance from first-order MMR for different measurement quantities using Q16 data (similar to Figure \ref{zetapic}).  Adjacent planets are shown with black dots while non-adjacent planets are gray dots.  The gray shaded regions indicate the boundaries for each MMR where each quantity is forbidden.  That is, a planet pair in the shaded region for one MMR is closer to a neighboring MMR and appears there.  The top panel shows the quantity $\Delta$ while the bottom panel shows the quantity $\zeta$.  We note that the deficit of planet pairs interior to first-order MMR is more prounounced using $\zeta$ (indeed, is more pronounced using Q16 than using Q8 data) and is nearly the same size for all indices $j$.  Notable planet pairs that lie in that gap are two non-adjacent pairs from KOI-730 near the 2:1, a pair from KOI-1599 near the 3:2, and a pair from KOI-430 near the 4:3.\label{zetapicq16}}
\end{figure*}

Plots of the quantities $\Delta$ and $\zeta$ for the Q16 candidates are shown in Figure \ref{zetapicq16}.  The dearth of planet pairs interior to first-order MMR is still visible (small positive values of $\zeta$), and indeed is more pronounced for MMRs of higher index.  Since the overall analysis of the raw \kepler\ data that produced the \citet{Burke:2014} catalog is not the same as the preliminary Q16 data, we expect that the list of planet candidates will continue to evolve as both the pipeline analysis and the candidate vetting process continue.

\section{Conclusions}\label{conclusions}

We have presented an analysis of the distribution of period ratios from \kepler\ planet candidates as identified in the Q8 catalog of \citet{Burke:2014}.  Our analysis is based upon a kernel density estimation of the true period ratio distribution using a Gaussian smoothing kernel.  The area under each gaussian is modified to account for the relative probabilities of observing each planet pair---correcting for both the geometrical effect of mutually inclined orbits and the effect of pipeline incompleteness.

For the geometrical effect, we assume that mutual orbital inclinations are Rayleigh distributed with a Rayleigh parameter of 1.5$^\circ$.  However, most of our analyses and conclusions (such as the statistical significance of certain features in the distribution) do not depend, or depend only weakly, on this assumption.  For example, if the height of the tail is too shallow because the true distribution is better modeled with a larger Rayleigh parameter, then the peak heights would grow, but so would the confidence regions from our Monte Carlo simulation, and their statistical significance would be largely unchanged.  Moreover, many of our claims regarding the relative importance of features in the period ratio distribution (such as the peak near 2.2) would actually be strengthened if orbital inclinations had larger variation.

We estimate the effects of pipeline completeness using the results from the TERRA pipeline \citet{Petigura:2012,Petigura:2013a,Petigura:2013b}.  We recognize that this is not identical to the \kepler\ pipeline---including the treatment of multiplanet systems and the magitude range of the stars considered (the injection and recovery tests of TERRA did not treat multiplanet systems and TERRA focused on the brightest \kepler\ targets).  However, while no direct comparison exists between the \kepler\ pipeline and TERRA (i.e. using identical raw data), the results in \citet{Petigura:2013a} indicate that most (more than 2/3) of the identified planets are common and that discrepancies between them can be largely explained by either planet multiplicity or by the fact that the TERRA pipeline was run on more data than was analyzed for the \citet{Batalha:2013} catalog---the catalog used for the comparison.  Still, the effects of pipeline completeness are significantly less than the effects of orbital geometry, and their inclusion has little material impact on our results, we believe that the TERRA pipeline serves as a valid benchmark for the efficiency of detecting planets in the \kepler\ data by an automated pipeline.

From our analysis of the distribution of period ratios we draw the following conclusions:
\begin{enumerate}
\item The previously identified features of a large excess of planet pairs near the 3:2 MMR and a large deficit of planet pairs just interior to the 2:1 MMR remain the largest excess and deficit respectively using data through Q8 (see Figures \ref{pdist} and \ref{pdistall}).
\item The large peak near the period ratio 2.2 is the second most prominent peak in the distribution (whether considering adjacent or all planet pairs).  A preliminary examination of Q16 \kepler\ data hints that this peak may ultimately be the most important feature in the distribution of period ratios---and by a substantial margin (cf., Figures \ref{mcpdist}, \ref{mcpdistall}, and \ref{q16pdist}).
\item In both the distribution for adjacent planets and the distribution for all planets we estimate with more than 90\% confidence that no more than one of the peaks and one of the troughs (and with 99\% confidence that no more than two peaks or troughts) that exceed the dotted curves in Figures \ref{mcpdist} and \ref{mcpdistall} are due to statistical fluctuations.  However, their origin may not be due directly to dynamics at the specified period ratio, but may be due to combinations of planet pairs with more common period ratios (see Figures \ref{newpeaks} and \ref{peak41}).
\item The overall peak of the period ratio distribution lies between the ratios of 1.5 and 2.0.  Beyond a period ratio of $\sim 2$, the probability density of all period ratios for \kepler -like planetary systems falls with a power-law exponent of $-1.26 \pm 0.05$ over at least a decade in period ratio and likely more (see Figure \ref{pdflines}).
\item When measured in terms of $\zeta$, the gap in planet pairs just interior to the 2:1 MMR appears for all values of the index $j$, but does not exist to the same degree when measured in the other terms shown in Equation \ref{quantities} (see Figure \ref{zetapic}).
\item While there are several planet pairs that are very near integer period ratios, we show that the number of these systems is consistent with randomly selected regions of identical size.  Thus, while high order $j$:1 resonant dynamics is not excluded in these systems, there is no evidence that orbits near these resonances are either favored or disfavored dynamically.
\item Finally, we identify several systems (Table \ref{closetable}) where resonant dynamical behavior is more likely to be found based upon the distance from first-order MMR measured in terms of the supremum of the libration width for each resonance (meaning that systems are likely no farther from the MMR than the values indicated).
\end{enumerate}

We believe that the overall properties of the period ratio distribution are important for our understanding of the formation and dynamical evolution of the inner regions of planetary systems.  Sharp features such as the peaks near 1.5 and 2.2 and the trough interior to 2.0 may indicate connections between the dynamics that we observe today and the interactions that the planets had while still embedded in the gas or planetesimal disks from whence they formed.  Broad features, such as the peak and tails of the distribution are also valuable for similar reasons.  We anticipate that future catalogs presenting \kepler\ planet candidates based upon all available data, as well as expected results from future transit surveys (e.g., NASA's Transiting Exoplanet Survey Satellite (TESS) and ESA's Planetary Transits and Oscillations of stars mission (PLATO)) and radial velocity surveys, will yield additional, important insights into the architectures, dynamics, and origins of planetary systems.

\section*{Acknowledgements}
We thank the Kepler mission for their hard work in making this mission a success and producing the catalogs used in this study.  We thank Erik Petigura for sharing the results of his work with us.  We also thank Sam Hadden, Eric Agol, Katherine Deck, Dan Fabrycky, Eric Ford, Jack Lissauer, Jerry Orosz, Darin Ragozzine, and William Welsh for useful discussions related to this work.  JHS is supported by a grant from the Kepler Participating Scientist Program (NNH12ZDA001N-KPS) and through the Lindheimer Fellowship at Northwestern University.  JAH acknowledges support from NASA through grant NNX12AI86G.  The authors are grateful to their host institution, the Center for Interdisciplinary Exploration and Research in Astrophysics (CIERA) at Northwestern University.

\appendix

\onecolumn

\section{Pipeline completeness correction}\label{apxpipecomplete}

\begin{figure}
\includegraphics[width=0.45\textwidth]{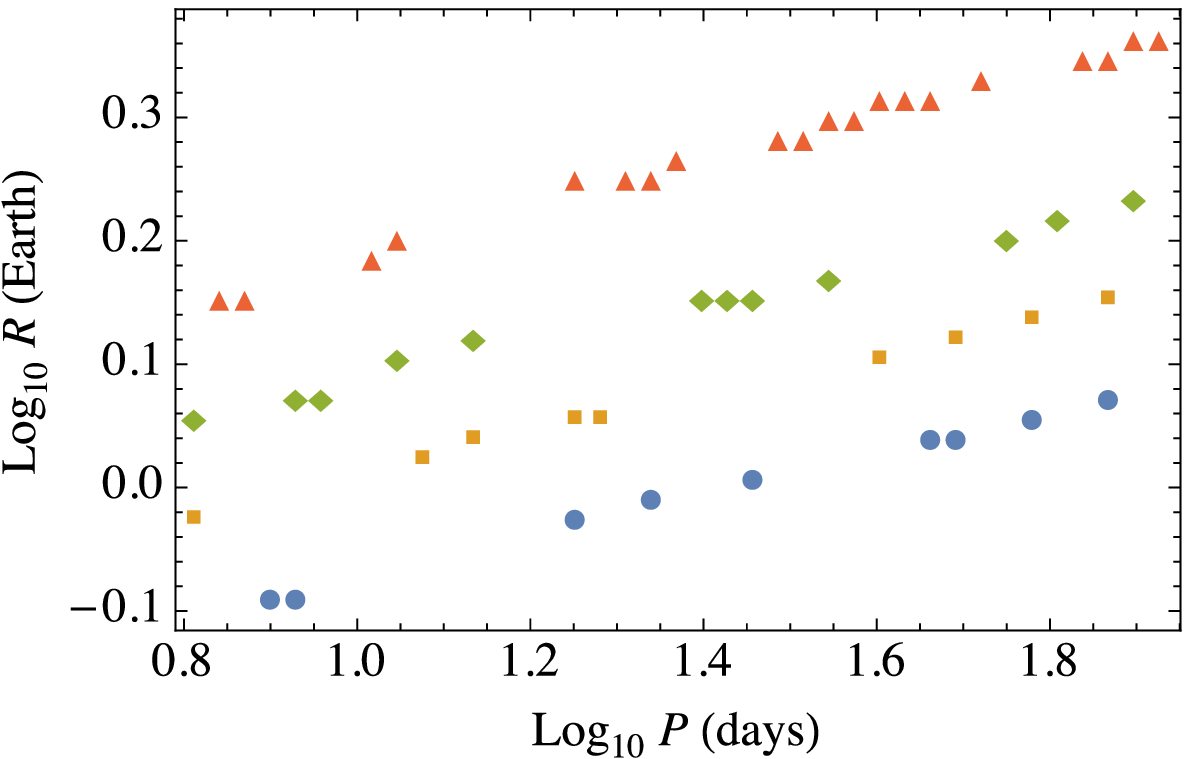}
\caption{Points from the TERRA analysis that lie within $\pm 0.005$ of the 0.2, 0.4, 0.6, and 0.8 completeness contours.  We remove the average slope of these lines (0.171) to produce a global completeness profile shown in Figure \ref{completeness2}.\label{compcontours}}
\end{figure}

We use the results of the completeness study for the TERRA pipeline using transit injection and recovery and shown in Figure 1 of \citet{Petigura:2013b}.  Those data were stored on a grid of completeness fraction as a function of orbtial period and planet radius.  Using data corresponding to orbital periods between $10^{0.8}$ and $10^2$ days we first select data where the completeness estimates lie within $\pm 0.005$ of 0.2, 0.4, 0.6, and 0.8---giving points along several lines of constant completeness.  These points are shown in Figure \ref{compcontours}.  We fit lines through the data for each of these contours (assigning equal weights to the points) and calculate the mean of the slope of those lines---0.171.  We use this slope as a means to rescale our semi-analytic completeness function.  A determination of the \kepler\ pipeline completeness and how it depends upon orbital period and planet size is ongoing \citet{Christiansen:2013} and will be useful for future studies similar to the one we conduct here.

\begin{figure}
\includegraphics[width=0.45\textwidth]{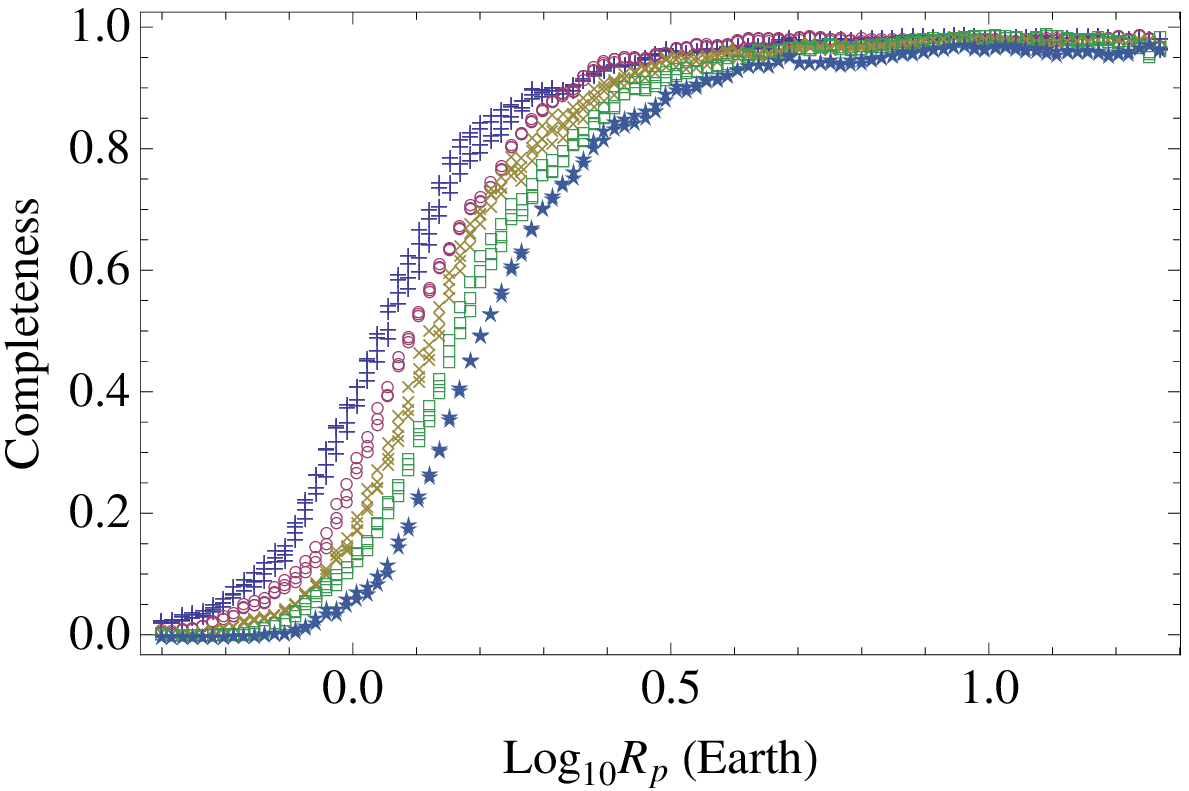}
\includegraphics[width=0.45\textwidth]{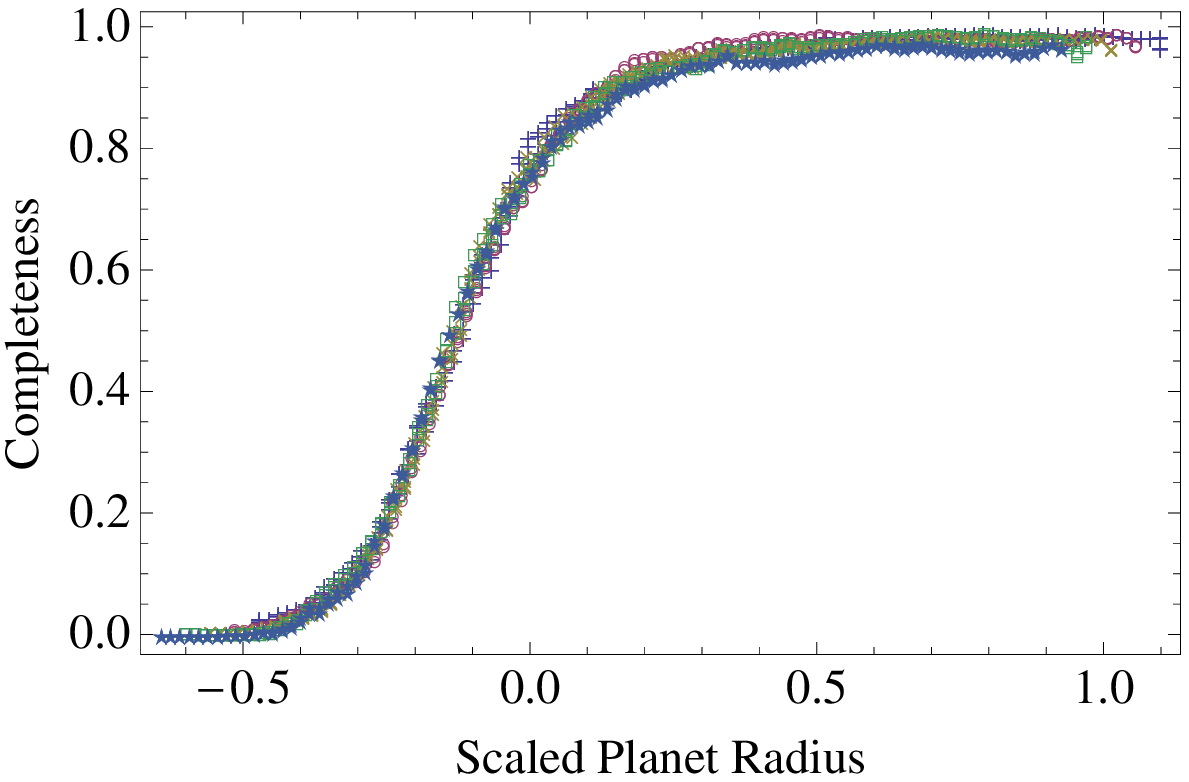}
\caption{The probability for the TERRA pipeline to recover planets as a function of planet radius.  The different sets of points correspond to different strips in orbital period ($10^{1.0}, \ 10^{1.25}, \ 10^{1.5}, \ 10^{1.75}, \ \text{and} \ 10^{2.0}$).  The left panel shows the raw results from \citet{Petigura:2013b}.  The right panel shows the results when a linear trend (in log space) is removed from the different curves.  This shows that the completeness function is separable in radius and period.  The results in the right panel are averaged and interpolated to produce the semi-analytic completeness correction shown in Figure~\ref{completeness}.\label{completeness2}}
\end{figure}

Next, we group the data into five narrow logarithmically-spaced bins in orbital period.  These stripes are $\pm 0.05$ wide (in log space) and are centered at values of 1.0, 1.25, 1.5, 1.75, and 2.0 (the outermost bin is slightly truncated, but this fact does not have any material effect on our results).  The completeness as a function of planet radius for these five stripes is shown in the first panel of Figure \ref{completeness2}.  We remove the linear trend identified in the previous paragraph and find that there is a roughly universal profile for the completeness as a function of planet radius---as seen in the second panel of Figure \ref{completeness2}.

Finally, we average the combined profile and interpolate using cubic spline interpolation to produce our pipeline completeness model shown as the straight lines in Figure \ref{completeness}.  We note that we do not model the sharp decline in completeness for orbital periods beyond a few hundred days, which occurs because of the limited time baseline of the observations.  Such a correction only affects about a dozen systems (depending upon the cut) and the majority ($\sim 75\%$) lie outside our region of primary interest with period ratios $\lesssim 5$.  Moreover, the data used to produce the \citet{Burke:2014} catalog is complicated by the fact that some \kepler\ diagnostics used data that extended beyond Quarter 8 of observations.

\section{Transit probability correction}\label{apxgeometry}

Here we derive the geometric correction we applied to the \kepler\ catalog to produce Figure \ref{geometry}.  In the calculations below we assume circular orbits.  We recognize that eccentricity can affect the probability of a planet transiting (e.g., \citet{Kipping:2014a}).  However, since we use these results to calculate ratio of the probability of detecting a planet at one orbital period with that same planet at a different period, the effects of eccentricity on the transit probability divide and have little effect on our results.

\subsection{Transit probability for a single planet}

A planet in a circular orbit transits if
\begin{equation}\label{EQ:ip0}
\sin(|i_\mathrm{p}|)\le \frac{R_\star}{a},
\end{equation}
where $i_\mathrm{p}$ is the projected inclination, $R_\star$ is the radius of the star, and $a$ is semi-major axis.  We define the z-axis, $\hat z$, towards the observer and the projected inclination, $i_\mathrm{p}$, as 
\begin{equation}\label{EQ:ip}
i_\mathrm{p} = \cos^{-1}(\hat z\cdot\hat l)-\frac{\pi}{2},
\end{equation}
where $\hat l$ is the normalized angular momentum vector,
\begin{equation}\label{EQ:lhat}
\hat l=R_y(\Omega)R_z(i)\hat y,
\end{equation}
$\Omega$ is the longitude of the ascending node, $i$ is the true inclination, and $R_y$ and $R_z$ are rotations by their arguments about the $y$ and $z$ axes respectively.

Combining \eqref{EQ:ip} with \eqref{EQ:lhat} we obtain
\begin{equation}\label{EQ:ip2}
\hat i_\mathrm{p} = \cos^{-1}(\sin(i)\sin(\Omega))-\frac{\pi}{2}.
\end{equation}
Finally, combining \eqref{EQ:ip0} with \eqref{EQ:ip2} we obtain the condition for observing a transit:
\begin{equation}
\sin(|i|)\le\frac{R_\star/a}{|\sin(\Omega)|}.
\end{equation}

We find the probability of transit for a single planet, assuming a circular orbit, $d=a$, by integrating between $0\le i\le 2\pi$ and $0\le\Omega\le\pi$.
\begin{equation}\label{EQ:Pgeneral}
P_\mathrm{transit}=\frac{1}{4\pi}\int^{\pi}_0\int^{2\pi}_{0}\Theta\left(\sin(|i|)\le\frac{R_\star/a}{|\sin(\Omega)|}\right)\sin(\Omega)\ \mathrm{d}i\ \mathrm{d}\Omega,
\end{equation}
where the function $\Theta(i)$ is a Heaviside function
\begin{equation}\Theta\left(\sin(|i|)\le\frac{R_\star/a}{|\sin(\Omega)|}\right)=
\begin{cases}
      1 & \sin(|i|)\le\frac{R_\star/a}{|\sin(\Omega)|}\\
      0 & \sin(|i|)>\frac{R_\star/a}{|\sin(\Omega)|}\\
\end{cases}.
\end{equation}

Evaluating the inner integral in \eqref{EQ:Pgeneral}, we find
\begin{equation}\label{EQ:fOmega}
\int_0^{2\pi}\Theta\left(\sin(|i|)\le\frac{R_\star/a}{|\sin(\Omega)|}\right)\mathrm{d}i=
\begin{cases}
      2\pi & 0\le\Omega\le\Omega_\mathrm{c}\\
      4\sin^{-1}\left(\frac{R_\star/a}{|\sin(\Omega)|}\right) & \Omega_\mathrm{c}<\Omega<\pi-\Omega_\mathrm{c} \\
      2\pi & \pi-\Omega_\mathrm{c}\le\Omega\le\pi\\
\end{cases},
\end{equation}
where $\Omega_{c} = |\sin^{-1}(R_\star/a)|<\pi/2$.  Inserting \eqref{EQ:fOmega} into \eqref{EQ:Pgeneral} yields
\begin{equation}\label{EQ:Ptransit}
P_\mathrm{transit}=\frac{1}{4\pi}\int^{\Omega_c}_02\pi\sin(\Omega)\ \mathrm{d}\Omega+\int^{\pi-\Omega_c}_{\Omega_c}4\sin^{-1}\left(\frac{R_\star/a}{|\sin(\Omega)|}\right)\sin(\Omega)\ \mathrm{d}\Omega+\int^\pi_{\pi-\Omega_c}2\pi\sin(\Omega)\ \mathrm{d}\Omega,
\end{equation}
which simplifies to
\begin{equation}\label{EQ:Ptransit2}
P_\mathrm{transit}=\frac{1}{\pi}\int^{\pi-\Omega_c}_{\Omega_c}\left(\frac{R_\star/a}{|\sin(\Omega)|}\right)\sin(\Omega)\ \mathrm{d}\Omega+(1-\cos(\Omega_c))
\end{equation}

\subsection{Transit probabilities for two planets}

We can find the probability of a specific subset of planets transiting by using \eqref{EQ:Pgeneral} and requiring that multiple conditions be fulfilled,
\begin{equation}\label{EQ:PMultiple}
P_\mathrm{transit}=\int^{\pi}_0\int^{2\pi}_{0}\prod\limits_{j=0}^\mathrm{n_t}\Theta\left(\sin(|i_j|)\le\frac{R_\star/a_j}{|\sin(\Omega)|}\right)\prod\limits_{k=0}^\mathrm{n_f}\Theta\left(\sin(|i_k|)\le\frac{R_\star/a_k}{|\sin(\Omega)|}\right)\sin(\Omega)\ \mathrm{d}i\ \mathrm{d}\Omega,
\end{equation}
where $j$ counts the indices over the planets that transit and $k$ counts the indices over the planets that do not transit. This expression may be evaluated numerically for each system.

For a two-planet system we define our reference plane to be aligned with one of the orbits and choose $\Omega=0$ to be where the orbital planes intersect. We can then find the probability of detecting a transit as a function of the mutual inclination, $\Delta i$, of the two orbits, where our reference plane is the orbit of the innermost planet and $\Delta i=i_1-i_2$.  The probability that both planets transit is then
\begin{equation}\label{EQ:P12}
P_\mathrm{12}=\int^{\pi}_0\int^{2\pi}_{0}\Theta\left(\sin(|i_1|)\le\frac{R_\star/a_1}{|\sin(\Omega)|}\right)\Theta\left(\sin(|i_1-\Delta i|)\le\frac{R_\star/a_2}{|\sin(\Omega)|}\right)\sin(\Omega)\ \mathrm{d}i\ \mathrm{d}\Omega.
\end{equation}

Using \eqref{EQ:Ptransit2} and \eqref{EQ:P12}, we find the probability of seeing both planets transit, assuming we see the inner planet transit and a Rayleigh distribution for the mutual inclination, as $P = P'_{12}/(P'_{12}+P'_1),$ where
\begin{equation}
P'=\int^{\pi/2}_0\frac{\Delta i}{\sigma^2}e^{\frac{-\Delta i^2}{2\sigma^2}}P\ \mathrm{d}\Delta i,
\end{equation}
and $\sigma$ is the standard deviation of the Rayleigh distribution.  We evaluate this equation at several values of $a/R_\star$ and $\prat$ to produce the transit probabilities shown in Figure \ref{geometry}.

\clearpage

\twocolumn

\bibliographystyle{plainnat}
\bibliography{multis}

\bsp

\label{lastpage}

\end{document}